%% file: main.tex
\documentclass[conference]{IEEEtran}

\usepackage{microtype} 
\usepackage[mathletters]{ucs}
\usepackage[utf8x]{inputenc}
\usepackage[english]{babel} 

\usepackage{amsmath}

\usepackage{hyperref} 
\usepackage{graphicx}
\usepackage{booktabs}

\usepackage{caption}
\usepackage{subcaption}
\usepackage{amssymb}
\usepackage{cleveref}
\usepackage{todonotes}

\usepackage{algorithm}
\usepackage{algpseudocode}

\usepackage{minted} 
\input{listings/output/pygmentsDefs}
\expandafter\def\csname PY@tok@err\endcsname{} 

\expandafter\def\csname PY@tok@go\endcsname{\def\PY@tc##1{\textcolor[rgb]{0.53,0.53,0.53}{##1}}}

\newcommand{\ua}{u_\alpha}
\newcommand{\ub}{u_\beta}
\newcommand{\ug}{u_\gamma}
 
\newcommand{\ca}{c_{q\alpha}}
\newcommand{\cb}{c_{q\beta}}
\newcommand{\cg}{c_{q\gamma}}

\newcommand{\bxi}[0]{\pmb{\xi}}
\newcommand{\bu}[0]{\pmb{u}}

\newcommand{\pystencils}{{\em pystencils}}
\newcommand{\lbmpy}{{{\em lbmpy}}}
\newcommand{\walberla}{\textsc{waLBerla}}

\title{lbmpy: Automatic code generation for efficient parallel lattice Boltzmann methods}

\begin{document}

\author{
\IEEEauthorblockN{Martin Bauer}
\IEEEauthorblockA{\textit{Chair for System Simulation,}\\
\textit{FAU Erlangen-Nürnberg} \\
	martin.bauer@fau.de}
\and
\IEEEauthorblockN{Harald Köstler}
\IEEEauthorblockA{\textit{Chair for System Simulation,}\\
\textit{FAU Erlangen-Nürnberg} \\
	harald.koestler@fau.de}
\and
\IEEEauthorblockN{Ulrich Rüde}
\IEEEauthorblockA{\textit{Chair for System Simulation,}\\
\textit{FAU Erlangen-Nürnberg} \\
\textit{CERFACS, 31057 Toulouse Cedex 1, France} \\
ulrich.ruede@fau.de}}

\maketitle


\begin{abstract} 
Lattice Boltzmann methods are a popular mesoscopic alternative to classical computational fluid dynamics
based on the macroscopic equations of continuum mechanics.
Many variants of lattice Boltzmann methods have been developed
that vary in complexity, accuracy, and computational cost. 
Extensions are available to simulate multi-phase, multi-component, turbulent, and non-Newtonian flows. 
In this work we present \lbmpy{}, a code generation package that supports a wide variety of different 
lattice Boltzmann methods.
Additionally,  \lbmpy{} provides a generic development environment for new schemes.
A high-level domain-specific language allows the user to formulate,
extend and test various lattice Boltzmann methods.
In all cases, the lattice Boltzmann method can be specified in symbolic form. 
Transformations that operate on this 
symbolic representation yield highly efficient 
compute kernels.
This is achieved by automatically parallelizing the methods, and
by various application-specific automatized steps that optimize the resulting code.
This pipeline of transformations can be applied to a wide range of lattice Boltzmann variants,
including single- and two-relaxation-time schemes, multi-relaxation-time methods, 
as well as the more advanced cumulant methods, and entropically stabilized methods. 
\lbmpy{} can be integrated into high-performance computing frameworks
to enable massively parallel, distributed simulations.
This is demonstrated using the \walberla{} multiphysics package
to conduct scaling experiments on the SuperMUC-NG supercomputing system
on up to 147\;456 compute cores.
\end{abstract}


\section{Introduction}
Computational science and engineering is an interdisciplinary field.
The workflow of creating computational models starts at (physical) reality and ends
with the production of efficiently executable computer code \cite{ruede2018research}.
However, on the route from physical phenomena to machine code  
lies the formulation of mathematical models and their discretization,  
the construction of time stepping schemes and solution methods,
the design and analysis of parallel algorithms,
the realization of complex software systems,
and finally the transformation to code that can be executed
on a given hardware.
The target computer architecture may be a
massively parallel system, possibly heterogeneous and
using accelerators that can only be exploited by special programming techniques.
During the development, many choices must be taken and alternatives considered.
Thus, creating computational science software is a work-intensive,
time-consuming, and error prone task,
whose complexity is easily underestimated, despite its fundamental relevance 
for extracting reliable predictions from scientific principles.
In this article, we will present progress towards the systematic design of scientific software based on 
the automatic derivation of methods including the automatic design of efficient software.

When designing scientific software with conventional programming techniques, it is
often difficult to find the right balance between the flexibility of an approach
and its performance, since these
are often conflicting goals.
Often the specialization to a restricted class of problems would permit special optimizations
and can thus lead to more efficient codes.
However, a flexible software design may lead to more extensible and more generally usable software.
Additionally, using general-purpose libraries for subtasks, such as for the solution of linear systems,
may reduce development time, but may also lead to overheads when special structures that would lead to faster algorithms can not be exploited. 

Furthermore, it is typical that computational software outlives the computer systems that it was originally designed for.
This leads to the problem of performance portability.
The choice of a specific algorithm and a specific software design
may have been good for the efficient execution on older computer architectures, 
but these design choices may turn out to be
a major bottleneck on the accelerator-based architectures that
dominate high-end computing today.
A complete rewrite would be necessary, but since this is too time-consuming and expensive.
Legacy code today may stay in use though it severely underperforms on modern hardware.

A generic approach to overcome these difficulties are
more advanced abstractions.
For example, libraries such as Kokkos~\cite{edwards2014kokkos} 
or programming systems such as OpenACC~\cite{farber2016parallel} 
present abstractions of fine-granular concurrent execution on modern hardware that
can help to alleviate the problems of performance portability.
An alternative to such libraries can be metaprogramming techniques and the usage of  
domain specific languages (DSL).
Here machine optimized code is generated utilizing 
program transformations and compiler technology. 
Using DSLs opens additional possibilities based on application-specific abstractions.
A prominent and successful example for a finite element specific DSL is UFL (unified form language)~\cite{alnaes2014unified} which is embedded in python. It is e.g. used in FeniCS \cite{logg2012automated} or Firedrake~\cite{rathgeber2016firedrake}. 
These automated computing platforms permit the generation of computational models based on
a wide variety of partial differential equations that can be expressed in a DSL.
For stencil based computations there exist a number of different DSLs e.g.~\cite{tang2011pochoir,holewinski2012high,henretty2013stencil,gysi2015stella} that work on structured grids and e.g. the ExaSlang DSL that can also work on block-structured grids \cite{lengauer2020exastencils}. 
These DSLs succeed in supporting a wide variety of mathematical models with high flexibility.
Additionally, most of them also support parallel program execution.
Such approaches can exploit application knowledge represented in the DSL design 
and can thus use optimization techniques
that are more powerful than optimizing compilers for general-purpose languages.

In the present article, we will focus alternatively on the lattice Boltzmann method (LBM), 
as a promising mesoscopic modeling paradigm.
Our scope of modeling thus supports kinetic schemes
as an alternative to continuum mechanics for fluid dynamics.
Kinetic schemes are often performance-hungry, but they also offer a high degree of parallelism.
Therefore the scalability and performance on modern computer architectures is a central goal
of our work.
In particular, we will attempt to reach the performance of the best manually optimized LBM  codes
by developing a new application-specific code generation technology. 
To achieve this, we will leverage long term experience in manually optimizing 
general stencil codes 
and systematic performance engineering for LBM methods
\cite{wilke2003cache,wellein2006towards,Wellein2006,godenschwager2013framework,feichtinger2015performance}.
In particular, we will build on \walberla{} as a state-of-the-art LBM software framework with excellent
performance characteristics \cite{bauer2020walberla,liu2019sunwaylb}.

However, the current article goes beyond developing tools that help to optimize given kinetic simulation algorithms for a variety of computer architectures. 
While this alone is also useful,
we here extend the code transformation approach to the earlier stages of the
computational science workflow. 
In particular, we point out that the design and derivation of lattice Boltzmann models follows
a complex but systematic methodology.
The development of a specific LBM for a given application
is characterized by various options and choices that the method designer must make.
This involves selecting a lattice model, defining momenta and relaxation rates, and many more,
as will be elaborated in detail below.
Based on these choices, a very wide variety of LBMs can be derived.
While any such model could be constructed manually and then optimized using
program generation technology, this article goes an essential step further.
The manual development of advanced LBM can also be time-consuming and error-prone.
We will present here, how the derivation of the models themselves can be conceptualized so
that it becomes amenable to automation.
In \lbmpy{}, the tedious mechanical steps of the LBM development
can be performed by automatic symbolic manipulations,
saving precious developer time and making the methods more reliable
and reproducible.

In summary, \lbmpy{} jointly with \walberla{} becomes a computing platform for LBMs that is
equivalent to what FEniCS is for finite elements.
\lbmpy{} helps to realize highly efficient LBM implementations
and makes it easy for the developer to experiment with different variants of the methods.
Fully grown implementations of many different LBMs can be generated with a single mouse click.
We emphasize here specifically that our code generation technology has the unique 
capability to generate highly efficient 
parallel code that is ready to run with optimal scalability on the largest supercomputers.

The LBM is based on concepts from statistical mechanics.
The fluid is modeled on the mesoscopic level using distribution functions that represent the statistical behavior of the particles constituting the fluid. 
Compared to traditional computational fluid dynamics (CFD), 
that describes the fluid macroscopically with the Navier-Stokes equations,
the mesoscopic description 
permits higher modeling expressivity, leading to a variety of LBMs,
e.g., for porous media or multi-phase flows. 
Its local data access pattern makes the method very well suited for modern hardware architectures that draw their computational power of ever-increasing concurrency.
To run code efficiently on these architectures, parallelism on different levels 
must be exploited, starting from single-instruction-multiple-data (SIMD) instruction sets,
utilizing multiple cores per node with OpenMP, 
to distributed memory parallelism with MPI.
Optimizing LB compute kernels to get the best possible performance requires hardware-specific adaptation.
This process costs significant development time.
Unfortunately, it must be repeated for each new hardware platform. 
Furthermore, the optimization process often
leads to code that is hard to read and maintain. 
In practice, this can lead to the effect that for prototyping and development, a slow but flexible code is used, and only a few proven methods are re-implemented in a highly optimized way.

Over time a large variety of LBMs have been proposed~\cite{Kruger2016}.
Starting from the simple, but widely used, BGK single-relaxation-time (SRT) operator that relaxes the current state linearly to equilibrium with a single relaxation rate, over two-relaxation-time methods (TRT)~\cite{Ginzburg2008} to general multi-relaxation-time (MRT)~\cite{DHumieres2002} methods.
All these methods relax to equilibrium in moment space and can be viewed as special cases of MRT.
Then, there are multiple advanced methods like cumulant~\cite{Geier2015} or entropic LBMs~\cite{Bosch2015}, where a different set of statistical quantities is chosen for relaxation or the discrete equilibrium is modified. 
All LB versions have in common that they are parameterized by a set of relaxation rates, which can either be chosen constant or adapted locally. 
Computing relaxation parameters from local quantities, e.g., shear rates, is used to implement turbulence models or to simulate non-Newtonian fluids.
Entropic KBC-type models~\cite{Bosch2015,Ansumali2003,Frapolli2017} can also be placed into this group since they are constructed based on an MRT method with two relaxation rates, where one relaxation rate is chosen subject to a local entropy condition.
Additionally, LBMs could be extended with different forcing schemes~\cite{silva2012first,guo2002discrete}, and the equilibrium could be approximated up to different orders either in the so-called incompressible~\cite{He1997a} or the standard compressible version.
During the implementation phase, the space of options is growing still. 
Different storage patterns for the distribution functions can be chosen~\cite{aapattern,Geier2017}. 
Hardware-dependent optimizations, like loop splitting and non-temporal stores, further increase the performance on recent CPU architectures.

Faced with this space of physical models and implementation/optimization options,
developers of LB frameworks are faced with two options: 
Either pick a small set of LB methods and optimize them for a specific target architecture,
repeating the full process when a new LB method or hardware platform
must be supported or, trying to abstract and automate the development task.
In this work, we show that it is indeed possible to automate many tedious development tasks,
like reformulating equations to save floating-point operations, 
splitting loops for better memory access behavior, or fusing stream and collision kernels.
We present the lattice Boltzmann code generation package \lbmpy{} that solves these problems by automating large parts of the LBM development and optimization process. Its source code and documentation are available open-source under the GNU AGPLv3 license~\footnote{\url{https://i10git.cs.fau.de/pycodegen/lbmpy}}.

\lbmpy{} is a system for development of computational fluid dynamics codes based on the LBM.
In this regard, it is comparable to other LBM frameworks such as OpenLB~\cite{Heuveline2007,OpenLB}, Palabos~\cite{Lagrava2012,Palabos}, elbe~\cite{Mierke2018,elbe}, LB3D~\cite{Groen2011,schmieschek2017,LB3D},
and HemeLB~\cite{Groen2013}.
Some of these LB frameworks like Sailfish~\cite{Sailfish} and TCLB~\cite{TCLB}
also use metaprogramming techniques, mainly for portability to GPUs.  
While these frameworks support many LB methods, systematic performance evaluations and
optimizations are predominantly available for simple LB collision operators like single- and two-relaxation-time methods~\cite{Wittmann2018,Wittmann2013a,Wellein2006,godenschwager2013framework}.

In this work, we first give an overview over the  specific LBM design workflow and the
associated code generation pipeline in~\cref{sec:pipeline}.
Then we describe the formalism for LB method specification in \cref{sec:modeldescription} that comprises the two upper abstraction layers. The transformation to an algorithmic description and its optimization is discussed in \cref{sec:kernelgeneration}. Finally, we present performance and scaling results in \cref{sec:performance}.

\section{Overview: LB Metaprogramming Pipeline}
\label{sec:pipeline}

In this section, we first give an overview of the various abstraction layers of \lbmpy{}'s metaprogramming pipeline. In the following sections, we then discuss each layer in detail. 
All abstraction layers are implemented based on the computer algebra system {\em sympy}. 
The code generation system itself is implemented in Python, 
but the generated code is produced in C/C++/LLVM for CPUs and in CUDA or OpenCL for GPUs. 

\begin{figure}[h]
   \centering 
    \includegraphics[width=\columnwidth]{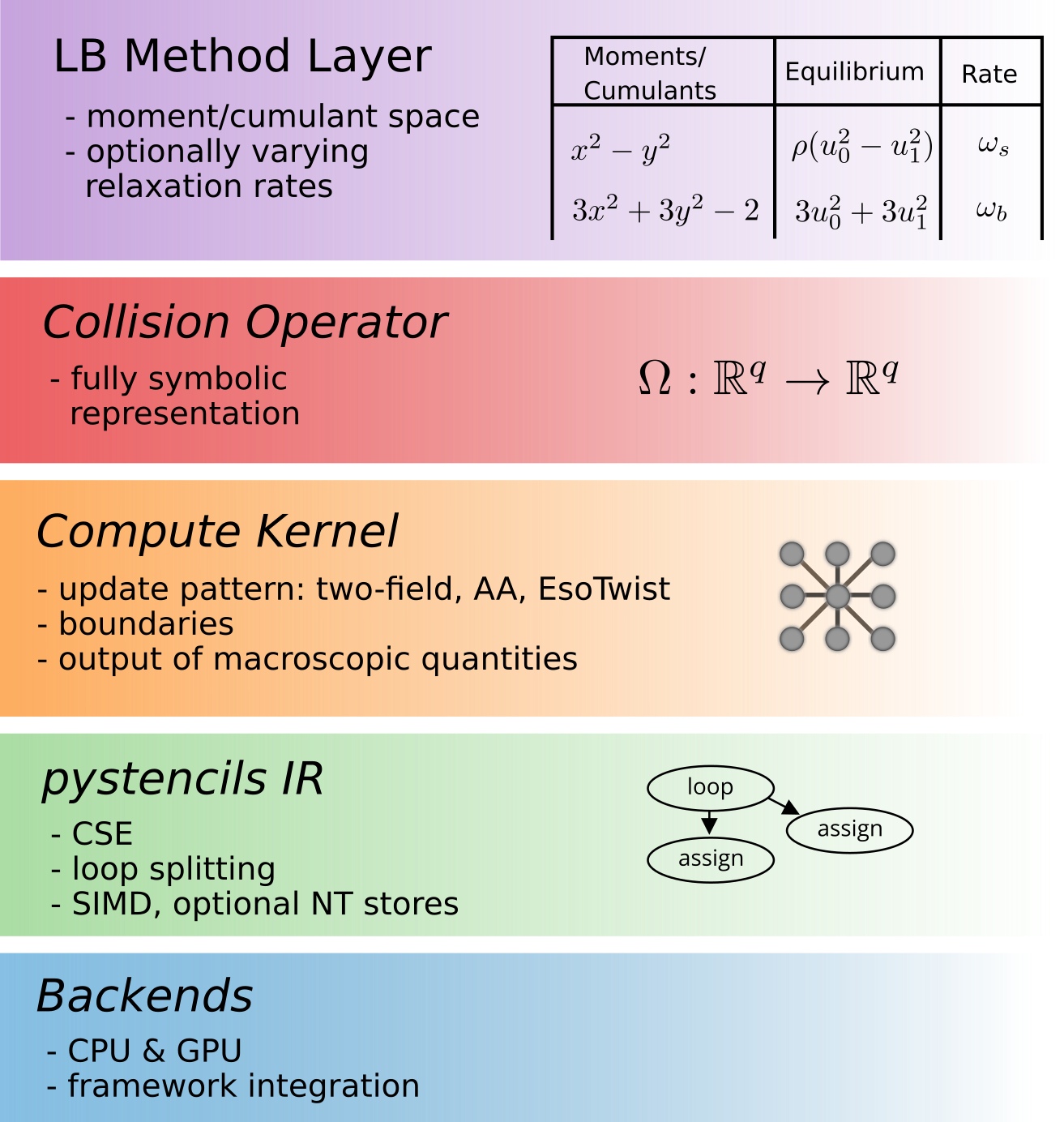}
    \caption{Abstraction layers of metaprogramming pipeline.}
    \label{fig:abstraction_layers} 
\end{figure}

As illustrated in~\cref{fig:abstraction_layers} the most abstract layer represents a LBM in $q$-dimensional collision space for a DdQq stencil. This collision space is either based on moments or cumulants \cite{Geier2015}.
To specify an LB scheme, the user defines a basis of the collision space as well as an equilibrium value and a relaxation parameter for each component. 
Relaxation parameters do not have to be constants but can be chosen as a symbolic expression that depends on local quantities like shear rates.
This permits the formulation of models for non-Newtonian fluids, turbulence models, or entropic stabilization.

\lbmpy{} transforms this high-level representation into a symbolic description of the collision operator. The collision operator is stored as a symbolic function $\Omega: \mathbb{R}^q \rightarrow \mathbb{R}^q$, where $q$ is the number of discrete distribution functions.

In the next step, the collision operator is mapped to a computational kernel.
This stage allows the generation of pure collision kernels as well as fused stream-collide kernels.

Additionally, a concrete storage pattern for the distribution functions is selected, e.g., two-array push or pull patterns or more elaborate single array storage schemes.
Optionally one can integrate boundary handling or output of macroscopic quantities at this layer as well.
The output of this stage is a symbolic stencil representation of an LB compute kernel. 
This stencil representation is passed to the \pystencils{} package~\cite{bauer2019code} that produces the actual code for CPUs or GPUs. \pystencils{} is extended with custom optimization passes to extend the code-generation pass with domain-specific knowledge.
In this work we focus on the CPU backend, GPU-specific optimizations and performance results are part of a second publication.

\section{Model description in collision space}
\label{sec:modeldescription}

In this section we first give a short overview over the theory that is used to define LBMs on the highest abstraction level.
Then we describe the ``LB method layer'' in more detail followed by examples showing how common collision operators can be specified on this layer. 

\subsection{Collision in moment space}

All LBMs considered here discretize the computational domain using a Cartesian, uniformly spaced grid.
The discretization is specified in stencil notation as $DdQq$, describing a $d$-dimensional domain where each cell contains $q$ particle distribution functions (PDFs) labeled $f_q(x_i, t)$ with $q \in \{1, ..., q \}$.
The PDF $f_q$ represents the mass fraction of particles moving along a lattice velocity $\mathbf{c}_q$.
In the following, we use lattice units, i.e., the positions $x_i$ and the time $t$ are integers.

The most basic and probably also the still most commonly used collision operator for
LBMs is the single-relaxation-time (SRT) or BGK operator. 
It relaxes each population to its equilibrium distribution using a single relaxation parameter $\omega$.
The SRT LBM can be succinctly written as
\begin{equation}
f_q(\mathbf{x} + \mathbf{c}_q, t + 1) 
= \omega \, f_q^{(eq)}(\mathbf{x}, t) + (1 - \omega) \, f_q(\mathbf{x}, t),
\label{eq:bgk}
\end{equation}
%
describing a single time step of the LBM, evolving the system from time $t$ to $t+1$. 
It can be split into a local collision step that computes a convex combination of the current state with the equilibrium state and a non-local streaming step that copies PDFs to neighboring cells.
The collision is formulated using a relaxation rate $\omega$ which is the inverse of the relaxation time $\tau$, i.e., $\omega = 1 / \tau$.
The local density $\rho$ and velocity $\mathbf{u}$ are computed from the distribution function as
\begin{equation}
\rho = \sum_q f_q  \hspace{0.5cm} \mbox{and} \hspace{0.5cm} u_i = \frac{1}{\rho_0} \sum_q c_{qi} f_q.
\label{eq:density}
\end{equation}
With these macroscopic values the equilibrium is given as
\begin{align}
f^{(eq)}_q =  w_q \rho + w_q \rho_0  
                     \big[  &\frac{\ca u_{\alpha}}{c_s^2} 
                              + \underbrace{  \frac{1}{2 c_s^4} \ua \ub (\ca \cb - c_s^2 \delta_{\alpha\beta})  }_{\mbox{2nd order}} \nonumber  \\
                             & + \underbrace{  \frac{1}{2 c_s^6} \left( \ua \ub \ug ( \ca \cb \cg - c_s^2 \ca \delta_{\beta\gamma} )  \right) }_{\mbox{3rd order}}
                     \big].
\label{eq:discrete_eq}
\end{align}
The reference density $\rho_0$ can either be chosen as $1$ to obtain a so-called incompressible LBM~\cite{He1997a}, for standard compressible LBM set $\rho_0=\rho$. Note that both versions approximate the incompressible Navier-Stokes equations (NSE)~\cite{Kruger2016}.
Equation \eqref{eq:discrete_eq} shows a third-order equilibrium approximation. To obtain the NSE in the macroscopic limit, the equilibrium is required only up to second order in $\mathbf{u}$~\cite{Kruger2016}. 

While the BGK collision operator is still widely used in practice, we aim for a more generic description of LBMs. 
To develop a high-level description of LBMs that is used as the input to the metaprogramming pipeline, we rely on the formalism of multi-relaxation-time methods. It includes the BGK operator and the popular two-relaxation-time method as special cases. 
With this approach \lbmpy{} is able to generate all LB schemes whose collision operator can be written in linear matrix form. This covers the majority of LBMs, with the most notable exception being the cumulant collision operator
that will be treated in \cref{sec:cumulant}.

The MRT formalism also allows us to derive the discrete equilibrium~\eqref{eq:discrete_eq} from its continuous counterpart instead of manually specifying it.
The collision operator of MRT methods first transforms the PDFs from population space into moment space via a moment matrix $\mathbf{M}$. The components of this moment vector $\mathbf{m} = \mathbf{M}\, \mathbf{f}$ are relaxed to equilibrium values $\mathbf{m}^{(eq)}$ using a diagonal matrix of relaxation rates $\mathbf{S}$. One collide-stream step of an MRT method then reads 
\begin{equation}
f_q(\mathbf{x} + \mathbf{c}_q, t + 1) 
= \mathbf{M}^{-1}\left[ \mathbf{S} \, \mathbf{m}^{(eq)}  + (\mathbf{I} - \mathbf{S})\, \mathbf{M} \, \mathbf{f} \right]
\label{eq:mrt}
\end{equation}
with $\mathbf{I}$ denoting the identity matrix. 
To fully specify the method, we have to 
define a concrete moment matrix $\mathbf{M}$, a vector of corresponding equilibrium values $\mathbf{m}^{(eq)}$, and a diagonal relaxation matrix $\mathbf{S}$.
We now discuss how each of these three ingredients can be specified in \lbmpy{}.

\subsubsection{Moment Space}
We begin with the transformation from population space to moment space via the moment matrix $\mathbf{M}$.
To derive an invertible transformation $\mathbf{M}$, a set of $q$ independent moments is required.
Moments can be identified with polynomials in the lattice velocities $\mathcal{P}(\mathbf{c_q}): \mathbb{R}^d\rightarrow \mathbb{R}$.
For example, the zeroth moment, i.e., the density, is given by the constant polynomial $1$. The first moments, i.e., the x, y, and z momentum densities, are represented by $c_{qx}$, $c_{qy}$, and $c_{qz}$.
Second order polynomials describe viscosity modes, for example in 2D, $c_{qx}^2 + c_{qy}^2$ is a mode related to bulk viscosity, whereas $c_{qx} \cdot c_{qy}$ and $c_{qx}^2 - c_{qy}^2$ are modes related to shear viscosity.
The moment value is then computed as
\begin{equation}
   \Pi_{\mathcal{P}}(\mathbf{f}) = \sum_{\mathbf{c_q} \in \mathcal{S}} \mathcal{P}(\mathbf{c}_q) f_q 
   \label{eq:disrete_moment_def}
\end{equation}
with a stencil $\mathcal{S}$ that is given by a sequence of directions with integer components.
Given a sequence of moment polynomials $\mathcal{P}_1 ... \mathcal{P}_k$ the elements of the moment matrix are computed as $M_{kq} = \mathcal{P}_k(\mathbf{c}_q) f_q$. 
So we have to select $q$ moment polynomials that yield an invertible moment matrix $\mathbf{M}$.
In \lbmpy{}, there are various options to provide these moment polynomials.
The most basic but also most flexible option is to list them explicitly. 
Then \lbmpy{} automatically computes the moment matrix and checks that it is invertible.
This option is useful if code for a given MRT method from literature has to be generated.

One central goal of \lbmpy{} is to derive LB methods automatically, and not having to pass in, for example, the discrete equilibrium or even the moment basis. 
Thus, our system additionally offers routines to construct the moment basis for first neighborhood stencils automatically.
First, these routines have to find $q$ monomials that lead to an invertible moment matrix $\mathbf{M}$, which can then be orthogonalized in a second step. 
To illustrate this procedure, we first consider the D3Q27 stencil. 
Since the velocity vector components only contain the values $\{-1, 0, 1\}$, moments with velocity powers larger than $2$ alias a lower order moment. For example 
\begin{equation}
 \sum_{\mathbf{c}_q \in \mathcal{S}} c_{qi}^3 f_q = \sum_{\mathbf{c}_q \in \mathcal{S}} c_{qi} f_q  \hspace{0.5cm} \mbox{if} \hspace{0.1cm} c_{qi} \in \{ -1, 0, 1\}.
\end{equation}
Similarly, moments with even exponents larger or equal than 4 are aliases by corresponding moments with exponent 2.
Potentially non-aliased moments are thus $c_{q0}^{e_0} c_{q1}^{e_1} c_{q2}^{e_2}$ with $e_i \in \{0, 1, 2\}$. 
These systematically constructed $27$ monomial moments can be used for the D3Q27 stencil to construct an invertible moment matrix. Similarly, in 2D, this strategy also yields an invertible moment matrix for the D2Q9 stencil. 
For D3Q15 and D3Q19 the situation is slightly more complex. 
For D3Q19, we start with the $27$ possible monomial moments and discard moments that produce a zero line in the moment matrix, like e.g. the moments defined by the polynomials $c_{q0}^2 c_{q1} c_{q2}$ or $c_{q0} c_{q1} c_{q2}$. For this stencil, there are in total $8$ out of the $27$ possible monomial moments leading to zero lines, leaving $19$ independent rows. 
For the D3Q15 stencil this procedure has to be further refined. There, some monomial moments produce the same non-zero row in the moment matrix, which are trivially linear dependent. 
\lbmpy{} groups moments together that yield the same row, resulting in 15 groups. 
One group of moments, for example is $[c_{q0}^2 c_{q1}, c_{q1}c_{q2}^2, c_{q0}^2 c_{q1} c_{q2}^2]$.
In each group, we keep only the lowest order moments. If there is more than one moment remaining, their sum is used. In case of above example this leads to $c_{q0}^2 c_{q1} + c_{q1}c_{q2}^2$. This systematic procedure constructs moment matrices that span the same space as MRT matrices reported in~\cite{DHumieres2002,Dunweg2007,Schiller2008}.

However, the constructed $q$ independent moments are not orthogonal yet. 
For MRT methods, typically, an orthogonal moment set is required. 
Using a symbolic Gram-Schmidt procedure, \lbmpy{} can orthogonalize the moments, either utilizing the standard scalar product, or a scalar product weighted by the lattice weights.
The exact outcome of the Gram-Schmidt orthogonalization depends on the ordering of the non-orthogonal moments that are put in. For reproducible results \lbmpy{} sorts the input by moment order and within each order lexicographically.
Before the orthogonalization, also the second-order moments are manually split into bulk and shear part.

\subsubsection{Equilibrium State}
The second element necessary for the construction of an MRT method is the equilibrium.
It can either be given in population space \eqref{eq:discrete_eq} or directly as a vector of equilibrium moments $\mathbf{m}^{(eq)}$. 
In this mode, the user has full control over the equilibrium values and can create LBMs not only for the Navier-Stokes equations but also for other partial differential equations. 

Beyond this, our meta-programming approach attempts to derive as much as possible from a more general formulation.
Therefore, we provide functionality in \lbmpy{}
to derive equilibrium values for hydrodynamic LBMs automatically.
One way to obtain a hydrodynamic discrete equilibrium is to compute the
equilibrium moments directly from the continuous Maxwell-Boltzmann distribution
\begin{equation}
  f^{(MB)}(\rho, \bu, \bxi) = \frac{\rho}{(2\pi c_s^2)^{\frac{D}{2}} } \exp\left( -\frac{||\bxi-\bu||^2}{2 c_s^2} \right)
\label{eq:cont_maxwellian}
\end{equation}
with
\begin{equation}
\int P(\bxi) \, f^{(MB)}(\rho, \bu, \bxi) \; d\bxi.
\label{eq:cont_moment_integral}
\end{equation}
In this case, the user first chooses a value for the speed of sound $c_s$, typically $c_s=1/\sqrt{3}$ for first neighborhood stencils, then the integral \eqref{eq:cont_moment_integral} is evaluated symbolically with the help of \textit{sympy}.
The resulting continuous moments of the Maxwellian are used in the equilibrium moment vector 
$\mathbf{m}^{(eq)}$. Optionally the moments can be truncated to a given order in the macroscopic velocity $\mathbf{u}$. If the equilibrium is required in population space, it can be easily transformed using the assembled moment matrix with $\mathbf{M}^{-1}\, \mathbf{m}^{(eq)}$.
For the cartesian product stencils D2Q9 and D3Q27, this method yields exactly the standard equilibrium~\eqref{eq:discrete_eq}. For the D3Q15 and D3Q19 velocity sets, however, a different equilibrium is obtained. A comparison of the standard equilibrium and the equilibrium obtained with this moment-matching technique can be found in~\cite{BAUER2020109111}.
The standard equilibrium for D3Q15 and D3Q19, including the weights, can also be derived by \lbmpy{} using a Hermite projection of~\eqref{eq:cont_maxwellian}. 


\subsubsection{Relaxation rates}
The third building block to fully define the method are the relaxation parameters. In \lbmpy{}, the user can specify a relaxation rate separately for each of the previously selected moments. 
Each relaxation rate can either be a constant value, a symbol, or an arbitrary expression of local or neighboring values. 
In the simplest case, the relaxation rate is a compile-time constant and equal for all time steps and lattice cells. 
This case allows the computer algebra system to pre-evaluate and simplify expressions containing constants only,
thus leading to significant savings.
If the relaxation parameter is chosen as a symbol, it becomes a run-time parameter of the generated kernel function.
In this case it can be changed, e.g., in a configuration file of the final application without requiring re-compilation.
Of course this comes possibly at the cost of executing more FLOPs as compared to the pre-evaluation.
The third option, where the relaxation rate is given as a symbolic expression, gives the most modeling power and flexibility.
The expression may contain any local or neighboring quantities like equilibrium or non-equilibrium moments. This allows the formulation of a wide range of turbulence models, where the relaxation rate needs to be adapted depending on shear rates. Entropically stabilized schemes like the KBC-type models~\cite{Bosch2015} can also be described in this way. 
The relaxation rate expression may also contain values of other arrays, allowing for easy coupling of multiple LB schemes, e.g., for multiphase or thermal flows. 


\subsection{Collision in cumulant space}
\label{sec:cumulant}

Recently an alternative collision space has been proposed in~\cite{Geier2015}.
Before collision, cumulants of the distribution function are calculated that are relaxed against their respective equilibrium value.
Conceptually, cumulant collision operators are realized in~\lbmpy{} similar to collision operators in moment space. The user specifies a set of cumulants, together with relaxation rates. The cumulant equilibrium values are obtained from the continuous Maxwellian. This allows the formulation of cumulant methods not only for D3Q27 and D2Q9 but also for D3Q19 and D3Q15 stencils. 

Cumulants can be succinctly defined through the cumulant-generating function 
\begin{equation}
K(\pmb{\xi}) = \ln\left( \sum_{\mathbf{c}_q\in \mathcal{S}} f_q \exp(\mathbf{c_q} \cdot \pmb{\xi} ) \right).
\label{eq:cumulant_generating_function}
\end{equation}
The cumulants are computed by multi-differentiation of \eqref{eq:cumulant_generating_function} and evaluating the derivative at zero. 
For example, the ``bulk cumulant'', that we associated with the polynomical $c_{qx}^2 + c_{qy}^2$ is computed as 
\begin{equation}
\frac{\partial^2 K(\pmb{\xi})}{\partial \xi_0^2}\bigg\rvert_{\pmb{\xi}=0} + \frac{\partial^2 K(\pmb{\xi})}{\partial \xi_1^2}\bigg\rvert_{\pmb{\xi}=0}.
\end{equation}
Originally, we implemented the cumulant transformation
with this approach in \lbmpy{}, however, the resulting expressions get very elaborate, especially for large stencils.
{\em sympy}'s common subexpression evaluation capabilities then run an unpracticable long time manipulating
these expressions.
Thus we have developed an alternative multi-step transformation, 
where the populations are first transformed to moment space and then to cumulants.
The moments are intermediate quantities that serve as common subexpressions.
In \lbmpy{}, we use Fa\`a di Bruno's formula \cite{roman1980formula} to derive the transformation of raw moments to cumulants and vice versa.

 
\subsection{Collision Model Examples}

In this section, we demonstrate how LBMs can be formulated in \lbmpy{} by constructing collision operators of varying complexity.
 
\subsubsection{SRT, TRT} 

We start with the single- and two-relaxation-time collision operators. 
Even if these methods are typically not derived in moment space, we use the MRT formalism for these operators as well to not introduce special cases.
The challenge with this general approach is, however, that the simplification system needs to be able to reduce the resulting expressions to their short form. 

The following code example shows the definition of a D2Q9 TRT method. 
Stencils are represented by a tuple of discrete directions with integer components. Common stencils, like the D2Q9, can be obtained by their name. 
This stencil is used to construct a set of independent raw moments using the algorithm described above. 

\vspace{0.3cm} 
\input{listings/output/00-trt-example.py}
\vspace{0.3cm} 

In this example, the moment equilibrium values are computed from the continuous Maxwellian, 
and the relaxation rates are defined for each moment.
\lbmpy{} offers various classification functions for moments like the \texttt{is\_even\_moment} function, used here. 
Other functions can determine the order of a moment, or if it is related to shear or bulk viscosity. 
Putting these elements together, the method is fully defined and can be displayed to the user in a Jupyter notebook~\cite{jupyter} in tabular form, as shown below. 
\vspace{0.3cm}

\begin{center}
\begin{tabular}{ccc}
Moment & Equilibrium & Relaxation rate \\
\hline
$1$ & $\rho$ & $\omega_{e}$ \\
$x$ & $\rho u_{0}$ & $\omega_{o}$ \\
$y$ & $\rho u_{1}$ & $\omega_{o}$ \\
$x^{2}$ & $\rho u_{0}^{2} + \frac{\rho}{3}$ & $\omega_{e}$ \\
$y^{2}$ & $\rho u_{1}^{2} + \frac{\rho}{3}$ & $\omega_{e}$ \\
$x y$ & $\rho u_{0} u_{1}$ & $\omega_{e}$ \\
$x^{2} y$ & $\frac{\rho u_{1}}{3}$ & $\omega_{o}$ \\
$x y^{2}$ & $\frac{\rho u_{0}}{3}$ & $\omega_{o}$ \\
$x^{2} y^{2}$ & $\frac{\rho u_{0}^{2}}{3} + \frac{\rho u_{1}^{2}}{3} + \frac{\rho}{9}$ & $\omega_{e}$ \\
\end{tabular}
\end{center}
\vspace{0.2cm}
For better readability we denote moment polynomials using variables $x, y$ and $z$ instead of $c_{qx}, c_{qy}$ and $c_{qz}$.
Note that no explicit equilibrium formulation similar to \eqref{eq:discrete_eq} was necessary to construct this method. Only the stencil, the continuous Maxwellian, and a systematically constructed set of independent raw moments have been used to derive this method.

\subsubsection{MRT}

Next, we show how to construct a generic MRT method in \lbmpy{}.
We stick with the D2Q9 stencil to keep the listing of the method tableaus short. 
Similar to the TRT method above, we start with a set of independent raw moments. 
For MRT methods, the moments have to be orthogonalized.
In \lbmpy{} we provide an orthogonalization routine based on the Gram-Schmidt procedure.
This routine either uses the standard or a weighted scalar product. 
A common choice is to use a scalar product weighted with the lattice weights,
which we demonstrate in the code example below. 
If we want to control bulk and shear viscosities using different relaxation rates, 
the second-order moments must be modified before the orthogonalization. 
This is handled by the \texttt{split\_shear\_bulk\_moments} function. 
We pass in the list of all raw moments, containing the second-order moments $x^2, y^2$ and $xy$.
These are split into the bulk moment $x^2 + y^2$ and the remaining $xy$ and $x^2 - y^2$ moments.
In 3D, this function works analogously. 

\vspace{0.2cm} 
\input{listings/output/01-mrt-example.py}
\vspace{0.2cm} 

The Gram-Schmidt orthogonalization step then produces the moments listed in the first column of the following table. 
Then a list is constructed that defines the relaxation rate for each moment. 
Moments of order less than two are conserved and the relaxation rate can be chosen arbitrarily. In this example, the relaxation rate is set to zero for these moments.
Having split up the second order bulk and shear moments, we can pick separate relaxation rates $ω_0$ and $ω_1$ for these. In this example, we choose a common relaxation rate for the third- and fourth-order moments. 

\begin{center}
\begin{tabular}{ccc}
Moment & Equilibrium & Relaxation rate \\  \hline 
$1$ & $\rho$ & $0$  \\ 
$x$ & $u_{0}$ & $0$  \\ 
$y$ & $u_{1}$ & $0$  \\ 
$x^{2} - y^{2}$ & $u_{0}^{2} - u_{1}^{2}$ & $\omega_{0}$  \\ 
$x y$ & $u_{0} u_{1}$ & $\omega_{0}$  \\ 
$3 x^{2} + 3 y^{2} - 2$ & $3 u_{0}^{2} + 3 u_{1}^{2}$ & $\omega_{1}$  \\ 
$3 x^{2} y - y$ & $0$ & $\omega_{2}$  \\ 
$3 x y^{2} - x$ & $0$ & $\omega_{2}$  \\ 
$9 x^{2} y^{2} - 3 x^{2} - 3 y^{2} + 1$ & $0$ & $\omega_{3}$  \\ 
\end{tabular}
\end{center} 

Compared to the TRT example, we have done another modification here. 
The equilibrium moments are modified to yield a so-called incompressible equilibrium~\cite{He1997a}. The incompressible equilibrium moments are obtained by
writing them as polynomial in the velocity $\mathbf{u}$ and substituting $\rho=1$ in all terms that contain at least one velocity component, e.g., $\rho + \rho u_0 \rightarrow \rho + u_0$.

Having full information about an LB method in the form of the moment table, as shown above, enables us to analyze the method using a Chapman-Enskog procedure symbolically, as long as relaxation rates are chosen constant.
The primary input for this analysis are the moment equilibrium values. The automated analysis can show the user the approximated PDE as well as higher-order error terms.
Additionally, it can derive the connection between relaxation parameters and macroscopic parameters, e.g., viscosities.
The following snippet shows the analysis of the MRT method defined here.
\vspace{0.2cm} 
\input{listings/output/04-ce-analysis.ipy}
\vspace{0.2cm} 


\subsubsection{Boundary Conditions}

Similar to the collision operator, boundary conditions are also described in symbolic form.
Boundary conditions have to specify the value of a population that is streamed in from a boundary lattice cell.
Here is an example of a velocity-bounce-back boundary that models a moving wall.

\vspace{0.2cm} 
\input{listings/output/03-ubb.py}
\vspace{0.2cm} 

In the boundary definition, the user has access to the method definition, that offers properties like speed of sound or lattice weights. The lattice direction \texttt{c} is an integer vector pointing from the fluid to the boundary cell. With this information, an expression for the missing population is constructed. The population field \texttt{f} and macroscopic properties can be accessed using relative addressing, where the center is the fluid cell. 

Additional information, like in this example, the velocity of the moving wall, can be used.
This data can be supplied by various sources. In the simplest case, it is a compile-time constant value. It can also be an expression that depends on spatial coordinates, time step, local population values, or macroscopic quantities.
It can also be supplied at runtime.
In this case the data is read from a field or a sparse list data structure
that stores this information for every connected boundary cell. 
More details will be covered in the section on the algorithmic treatment of boundary conditions.
However, in all these cases, the boundary definition, as shown in the above example, does not change at all. The definition and implementation are strictly separated.

Boundary conditions are defined per lattice link, not by lattice cell. That means, for example, that for each link a different velocity can be prescribed. 

\subsubsection{Turbulence models}

Up to now, we have presented methods with constant relaxation rates. 
The modeling power of LBMs stems partially from the ability to vary
relaxation rates on a cell-by-cell basis, depending on local quantities. 
With this technique, one can for example, model non-Newtonian fluids or implement turbulence models. 
To provide this modeling power to the user, \lbmpy{} 
does not only allow for compile- and run-time constants as relaxation rates.
It can also take arbitrary expressions of neighboring distribution functions or macroscopic quantities as relaxation rates. 
We illustrate this for the example of a
Smagorinsky subgrid turbulence model.

This model adds an eddy viscosity $\nu_t$ to represent energy damping on unresolved scales~\cite{Hou1996}.
The eddy viscosity is calculated from the local strain rate tensor as
\begin{equation}
\nu_t = \underbrace{(C_S \Delta)^2 |S|}_{\nu_{t}}
\end{equation}
where $C_S$ is a constant and $\Delta$ is a filter length chosen as 1 in lattice coordinates.
$|S|= \sqrt{2 S_{ij} S_{ij}}$ is the Frobenius norm of the local strain rate tensor
\begin{equation}
S_{ij} = \frac{1}{2} \left( \partial_i u_j + \partial_j u_i \right) = 
- \frac{3 \omega}{2 \rho} \Pi^{(neq)}_{ij}.
\label{eq:smag_shear}
\end{equation}
This uses the fortunate property of LBMs
that the strain rate tensor can be computed from local quantities, only using the second non-equilibrium moment~\cite{Kruger2009}
\begin{equation}
\Pi^{(neq)}_{ij} = \sum_q c_{qi} c_{qj} \; \left(f_q - f_q^{(eq)} \right).
\end{equation}
Equation~\eqref{eq:smag_shear} contains the total relaxation rate $\omega$ which is computed from the total viscosity $\nu = \nu_0 + \nu_t$ which again depends on the eddy viscosity $\nu_t$ that we want to determine
\begin{equation}
 \omega = \frac{2}{6 C_S^2 |S| + 6 \nu_0 + 1}.
 \label{eq:smag_omega}
\end{equation}
Thus we have a system of two equations in
$ω$ and $|S|$ that we now like to solve for $ω$. 
Here it pays off that \lbmpy{} is based on the computer algebra system {\em sympy} where these steps can be 
performed automatically:

\vspace{0.2cm} 
\input{listings/output/02-smagorinsky.py}
\vspace{0.2cm} 

The resulting symbolic expression for the effective $ω$ value can be be used in all places where in previous examples a constant has been used.
Thus, one can construct MRT or cumulant methods where some or all relaxation rates vary locally,
using potentially different expressions for different relaxation rates. 
For brevity, we have shown here only the construction of a simple turbulence model. 
The possibility to employ arbitrary expressions as relaxation rates in \lbmpy{} can be used to realize
also more advanced turbulence models with only little programming effort.
Additionally, \lbmpy{} also comes with several pre-defined turbulence models.
Thus the user does not have to perform the steps outline above manually, if only a common
turbulence model is required.  

\subsubsection{Entropic KBC Models}

Another important class of models, where relaxation rates are varied locally, are entropic LB schemes. 
In this section, we show how entropic MRT methods, labeled KBC models by the authors in~\cite{Bosch2015}, are realized in \lbmpy{}.
The central idea of these methods is to maximize a discrete entropy measure $S$ of the post-collision state.
The free variable that is tuned to obtain maximum entropy is a relaxation rate associated with higher-order moments. 
Single relaxation time entropic methods also change the effective
viscosity by varying this single rate to maximize entropy. 
KBC models present an improvement by using two relaxation rates:
One rate for the shear moments called $ω_s$, and a second relaxation rate $ω_h$ that controls higher order moments.
Only $ω_h$ is changed according to the entropy condition. 
The shear relaxation rate $ω_s$ is not altered, and thus also the viscosity remains constant.
By choosing which moment is relaxed by which relaxation rate, one obtains different KBC variants, that are labeled by the authors as KBC-N1 up to KBC-N4. 

In the original work~\cite{Bosch2015}, the notion of mirror states and according relaxation parameters is used. 
Here we use a different notation that is closer to the formalism of MRT methods. 
We start with an arbitrary MRT method that uses two symbolic relaxation rates $ω_s$ and $ω_h$. 
The rate $ω_s$ must include the shear moments if shear viscosity should remain constant.
The collision operator in population space is then of the form
\begin{equation}
 f'_q = f_q - ω_s Δs_q - ω_h Δs_h,
 \label{eq:entropic_linear_update}
\end{equation}
where $f'_q$ is the post-collision state, $f_q$ the pre-collision state, and $Δs_q, Δh_q$ being the coefficients multiplying the relaxation rates. 
Then we need to maximize the entropy 
\begin{equation}
 S(\mathbf{f}') = - \sum_q f'_q(ω_h) \ln \left( \frac{f'_q(ω_h)}{f_q^{(eq)}} \right)
 \label{eq:discrete_entropy_postcoll}
\end{equation}
in every cell at every time step by varying $ω_h$.
Taking the first derivative of~\eqref{eq:discrete_entropy_postcoll} w.r.t. $ω_h$ we get the optimality condition
\begin{equation}
 \sum_q Δh \left[ \ln\left( \frac{f'_q(ω_h)}{f_q^{(eq)}} \right) + 1 \right] = 0.
\end{equation}
This condition could be solved numerically in every cell using Newton's method. 
However, in this case, a more efficient way can be devised.
We expect $f'$ to be close to $f^{(eq)}$ and approximate the logarithm around $1$ up to first order 
with $\ln(x) \approx x-1 $. t
The optimality condition then simplifies to
\begin{equation}
 \sum_q Δh \; \frac{f'_q(ω_h)}{f_q^{(eq)}} = 0.
\end{equation}
Inserting the post-collision value as $f'_q = f_q - ω_s Δs_q - ω_h Δs_h$, and introducing the entropic scalar product $\left<a, b \right>_E := \sum_q a_q b_q \left[ f_q^{(eq)}\right]^{-1}$, we can solve for $w_h$ and obtain
\begin{equation}
 ω_h = 1 + (1 - ω_s) \frac{\left< Δs, Δh\right>_E}{\left<Δh, Δh\right>_E}.
 \label{eq:entropic_direct_omega_eq}
\end{equation}
To obtain this result, one has to replace $f_q = f_q^{(eq)} + Δs_q + Δh_q$ and use $\sum_q Δh_q = 0$, which holds because of the mass conservation property of the collision operator. 
All steps leading to~\eqref{eq:entropic_direct_omega_eq}, are implemented using {\em sympy}, to obtain an automatic derivation of KBC methods from high-level principles. 

Using this technique, we can construct a wide range of entropically stabilized methods, not only for D3Q27 stencils as in~\cite{Bosch2015} but for D3Q19 and D3Q15 stencils as well.
Furthermore, we offer a more costly but also more general numerical maximization procedure for the post-collision state entropy, which is based on Newton's method. This can e.g. be used for cumulant methods where the update is not linear in the relaxation rates any more as in~\eqref{eq:entropic_linear_update}, but has the quadratic form $f'_q = f_q - a_1 ω_s - a_2 ω_s^2 - b_1 ω_h - b_2 ω_h^2$.

\section{Compute Kernel Generation}
\label{sec:kernelgeneration}

All steps described up to now produce a symbolic representation of the collision operator $\Omega: \mathbb{R}^q \rightarrow \mathbb{R}^q$. Together with the stencil, represented as a list of $q$ discrete velocities with integer components, an efficient LB compute kernel must be generated for various hardware platforms.
This process is discussed in the following section.

\subsection{Simplification}

To obtain an efficient formulation of the resulting compute kernel, the symbolic collision operator 
must be rewritten in a form where as few as possible floating point operations (FLOPs)
are required to compute post-collision values. 
This is a very challenging task, since the automatic operator derivation yields a highly inefficient formulation by default. 
Consider, for example, the case of an SRT model that is derived by transforming populations to moment space, relaxing with a single rate, and transforming back. 
The matrix products produce lengthy expressions,
that are mathematically equivalent to the usual SRT formulation but are now expressed using many more FLOPs.
With standard mathematical techniques, like expanding or factoring, terms can already be simplified considerably. 
However, the most significant reduction in the number of FLOPs is achieved with common subexpression elimination (CSE). 
General CSE algorithms implemented in computer algebra systems are not guaranteed to find the global optimum and have to rely on heuristics to find a reasonably good solution. 
These algorithms do not just identify common subtrees as it is typically done as a compiler optimization, but they also try to rewrite the expressions in a form where they have more common subtrees. 

For an illustration of the optimization possible in \lbmpy{} we present here results for the D3Q19 BGK method.
The first row of Table~\ref{tbl:flops_D3Q19_SRT} shows the number of FLOPs in the expressions as they 
are produced by the automatic derivation. 
In total, this initial automatically generated code version needs 1263 operations.
To reduce cost, we first employ the simplification and CSE capabilities of the {\em sympy}
computer algebra system directly. The results are labeled ``{\em Only CSE}'' in the table. 
This reduces the number of operations significantly, down to only 261. 
However, a manually optimized implementation of the BGK method developed 
by the authors requires only 204 FLOPs.
A value around 200 FLOPs is also reported by Wellein et al.~\cite{Wellein2006}. 
The default simplification and CSE of {\em sympy} thus is unable to produce code as good as manually tuned, 
since it needs about 30\% more FLOPs than the best solutions known.
We also tested the simplification capabilities of other computer algebra systems, including Maple and Mathematica
which also could not find simplifications competitive with hand-tuned code.
Therefore, it was necessary to develop a new set of custom transformations
to rewrite the equations before they are passed to the CSE function of {\em sympy}.
These transformations are listed in the order of application
in the lower part of table~\ref{tbl:flops_D3Q19_SRT} as ``{\em Custom}'' transformations.
\begin{table}
\begin{center}
\begin{tabular}{l|cccc}
                       & Additions & Muls & Divs & Total \\
\hline
{\em Only CSE:} & & & \\
\hspace{1em}initial                 & 686  & 574  & 3 & 1263  \\
\hspace{1em}sympy CSE               & 199  & 61   & 1 & \textbf{261}   \\
& & & \\
{\em Custom:} & & & \\
\hspace{1em}initial                     & 686  & 574  & 3 & 1263 \\
\hspace{1em}expand                      & 173  & 423  & 3 & 599  \\
\hspace{1em}quadratic velocity prod.    & 203  & 447  & 3 & 653  \\
\hspace{1em}expand                      & 179  & 423  & 3 & 605  \\
\hspace{1em}factor $ω$'s                & 179  & 305  & 3 & 487  \\
\hspace{1em}common quadratic term       & 131  & 161  & 3 & 295  \\
\hspace{1em}substitute existing subexpr.& 119  & 119  & 3 & 241  \\
\hspace{1em}sympy CSE                   & 119  & 73   & 1 & \textbf{193} \\
\end{tabular}
\end{center}
\caption{Detailed simplification results for compressible D3Q19 SRT}
\label{tbl:flops_D3Q19_SRT}
\end{table}
Next to the name of the transformation we display the number of FLOPs after the transformation has been applied. 
Some of these transformations use LBM application knowledge, e.g., they treat density and velocity symbols differently. 

We now study these transformations one by one and describe them in detail.
The initial formulation is first expanded, i.e., transformed into a sum of products using a function provided by {\em sympy}.
The next transformation called ``quadratic velocity products'' is specifically developed for LBM simplification.
It picks out mixed quadratic terms in macroscopic velocity components $u_i u_j$ and replaces them by $(e^2 - u_i^2 - u_j^2)/2$, with a new subexpression $e := u_i + u_j$.
This transformation may seem counterintuitive since it increases the number of FLOPs.
However, it helps the following transformations to obtain better results.
This is an example of a transformation that requires domain knowledge since
this replacement may only be applied to the velocity symbols.
Next, a standard expansion is performed. 
The previously introduced subexpression $e$ prevents this expansion from undoing the previous transformation.
The next transformation uses a generic {\em sympy} function to factor out relaxation rates.
The ``common quadratic term'' step introduces a subexpression that is obtained by taking the expression for the center point, setting all pre-collision values to zero, and relaxation rates to one.
For the TRT method this yields
$ρ - 3/2ρ (u_0^2 + u_1^2 + u_2^2)$.
After this transformation, already existing subexpressions like density and velocity are searched in the equations,
and finally, a CSE from {\em sympy} is performed.
With this custom simplification pipeline, we eventually arrive at
a kernel that costs only $193$ FLOPs.
Note that this is even slightly better than the previously known and carefully hand-optimized version.
The improvement compared to using only the generic simplification is $35\%$.
Note also, that the same automatic simplification and optimization steps can now be performed for
other LBM methods.

Table~\ref{tbl:flops_all_methods} shows the total number of FLOPs for a selection of  LB schemes that \lbmpy{}
can generate and optimize.
We compare methods that use the so-called compressible and incompressible
equilibrium~\cite{He1997a}.
We also compare SRT, TRT, MRT, and the SRT with Smagorinsky turbulence model.
The table does not display in detail which type of FLOPs each method is composed of. 
However, all methods only require additions and multiplications with the following exceptions:
All compressible models have one division by the density, and the Smagorinsky methods
additionally require two square root operations.
The results in the first column are obtained by using only the {\em sympy} CSE. 
The third column uses the custom simplification strategy introduced above. 
The second column also uses the custom simplification strategy, with a modified CSE step at the end.
In this CSE step, we first search for subexpressions in terms that update opposing lattice directions.
By construction, these contain terms that differ in sign only and are good common subexpression candidates.
This step is then again followed by a global CSE.
This approach is labeled ``direction CSE'' since lattice directions are taken into account. 
The lowest FLOP count is marked for each method in boldface.
 

 
\begin{table}
 \begin{center}
  \begin{tabular}{l|ccc}
     & Only CSE & \parbox{1.6cm}{\centering Custom with \\direction CSE} & \parbox{1.5cm}{\centering Custom, \\default CSE}       \vspace{0.07cm} \\
     \hline
        {\em D2Q9}  & & & \\
        compr. SRT  &  113  & \textbf{90}  & \textbf{90} \\
        incompr. SRT  &  107  & \textbf{75}  & \textbf{75} \\
        compr. Smag.  &  137  & \textbf{122}  & \textbf{122} \\
        compr. TRT  &  114  & 110  & \textbf{101} \\
        incompr. TRT  &  108  & 103  & \textbf{94} \\
        compr. MRT  &  \textbf{150}  & 349  & 317 \\
        compr. weighted MRT &  \textbf{153}  & 350  & 325 \\        
        \hline
        {\em D3Q19}  & & & \\
        compr. SRT  &  261  & \textbf{193}  & \textbf{193} \\
        incompr. SRT  &  252  & \textbf{162}  & \textbf{162} \\
        compr. Smag.  &  306  & \textbf{251}  & \textbf{251} \\
        compr. TRT  &  262  & 233  & \textbf{214} \\
        incompr. TRT  &  253  & 225  & \textbf{206} \\
        compr. MRT &  \textbf{444}  & 1098  & 962 \\
        compr. weighted  MRT&  \textbf{406}  & 947  & 903 \\        
        \hline 
        {\em D3Q27 } & & & \\        
        compr. SRT  &  444  & \textbf{293}  & 389 \\
        incompr. SRT  &  435  & \textbf{289}  & 346 \\
        compr. Smag.  &  510  & \textbf{370}  & \textbf{370} \\
        compr. TRT  &  446  & \textbf{379}  & 516 \\
        incompr. TRT  &  437  & \textbf{374}  & 482 \\  
        compr. MRT  &  \textbf{651}  & 3155  & 4054 \\
        compr. weighted MRT &  \textbf{786}  & 3290  & 3984 \\        
        \end{tabular}
 \end{center}
 \caption{Total number of FLOPs for different LB schemes. 
          The ``Only CSE'' columns runs only a CSE from {\em sympy}. 
          The ``Custom with direction CSE'' runs the custom simplification pipeline,
          then a PDF direction-aware CSE followed by a standard CSE. 
          ``Custom with default CSE'' is the similar, but without the direction-aware CSE.
         }
 \label{tbl:flops_all_methods}
\end{table}
Let us first discuss the results for all methods with one or two relaxation rates.
We see that for SRT and TRT methods the custom simplification pipeline consistently leads to better results.
It is generic enough to work not only for the SRT method it was designed for,
but leads to good results for TRT methods as well.
Also turbulence models built on top of these collision operators are simplified better by the custom strategy,
as shown in the table with the Smagorinsky example.
Whether the direction-aware CSE is beneficial depends on the stencil.
For the D3Q27, it gives the best or equal result across methods,
for D2Q9 and D3Q19 it is helpful only for the SRT operators. 

We also have chosen two example MRT methods. One, that uses the standard scalar product for moment orthogonalization, and one with moments that are orthogonal w.r.t. to the weighted scalar product.
Second order shear and bulk moments are relaxed with different rates,
and for each order larger than 2 a separate relaxation rate is chosen.
Table~\ref{tbl:flops_all_methods} shows that the custom simplification pipeline cannot handle MRT methods.
A straight application of CSE obtains much better results for all tested MRT methods, regardless of the stencil. 

Currently we employ these three simplification options for each method, and then automatically
select the best one. 
In the future we plan to also use machine learning techniques to optimize the
application order of transformations or for finding new transformations.


%
%

\subsection{Collision Operator to Stencil}

\subsubsection{Streaming and Collision}

After simplification we have the collision operator given as a function $\mathbb{R}^q \rightarrow \mathbb{R}^q$. It is represented by a list of $q$ symbolic expressions for the post-collision population values accompanied by a set of subexpressions. 
The next stage transforms this formulation into a stencil representation. 

The stencil representation and all following low-level transformations are part of the {\em pystencils} package\footnote{\url{https://i10git.cs.fau.de/pycodegen/pystencils}} that is also developed by the authors~\cite{bauer2019code}. 
{\em pystencils} generates stencil kernels, i.e., routines
that iterate over arrays, applying the same operation on every cell.
It distinguishes between spatial and index dimensions.
Only spatial dimensions are iterated over, while index dimensions are used to address values stored inside a cell,
e.g., the $q$ populations for a PDF array or components of a vector field.
The central concept of {\em pystencils} are fields, and field accesses.
A field is defined by a name and its number of spatial and index coordinates.
Fields are indexed relatively, so the field access \texttt{f[1][0](q)}, for example, refers to the $q$'th population value of the east neighbor cell in a 2D setup. 
{\em pystencils} is built on top of {\em SymPy}, and field accesses can be used just like a built-in symbol. 
The collision operator can be transformed into a stencil representation by
replacing the pre- and post-collision symbols by field accesses.
Two additional pieces of information are required for this process.
The user has to choose the data layout of the population array and a kernel type that describes the operations done inside a kernel call.

\begin{figure}
 \begin{center}
   \includegraphics[width=0.7\columnwidth]{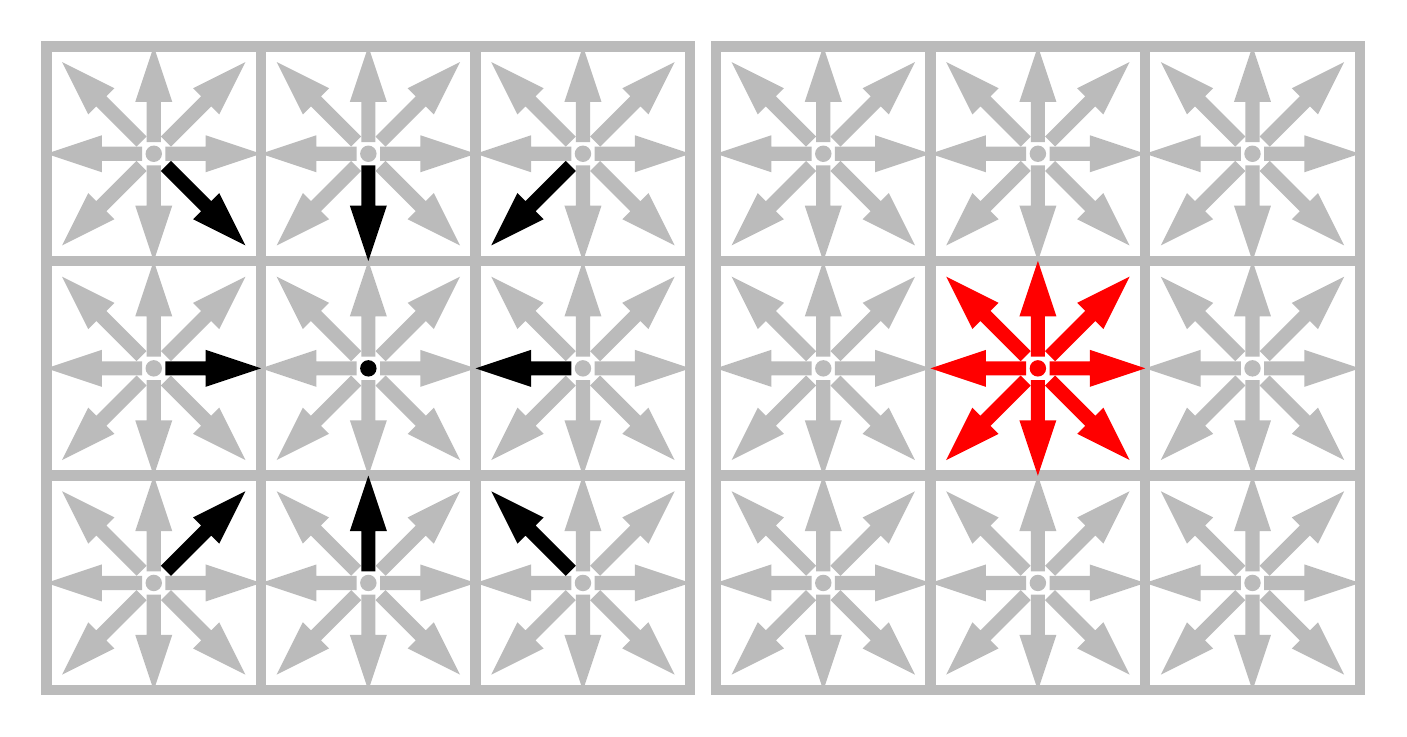} 
 \end{center}
 \caption{Visualization of stream-pull-collide update pattern using two arrays. The left part encodes the reads of pre-collision values, the right part shows where post-collision values are written to.}
 \label{fig:stream_collide}
\end{figure}

\lbmpy{} supports three different population storage options. The simplest approach is to have two arrays, where, during one kernel invocation, one array is read-only and the second array is write-only.
For this storage pattern the system can generate a fused stream-pull-collide, a fused collide-stream-push, or a collision-only kernel.
For a pure LBM simulation that is not coupled to other simulation models, typically a stream-pull-collide kernel gives the best performance.
In \lbmpy{} the data access patterns are encoded by the field accesses where pre-collision values are loaded from, and field accesses where post-collision values are written to.
These are visualized in~\cref{fig:stream_collide} for a stream-pull-collide kernel.
This mechanism cleanly separates the LB method definition from algorithmic- and data structure aspects, avoiding any code duplication.

Besides the simple two array swapping technique,
\lbmpy{} supports also more advanced storage patterns that operate on a single PDF array and thus require only half the memory. 
Supported single-array schemes are the AA pattern~\cite{aapattern} and the esoteric twist (EsoTwist) update scheme~\cite{Geier2017}. Single array storage schemes that introduce a data dependency between cell updates like~\cite{pohl2003optimization} are not supported.


\begin{figure}[H]
 \begin{center}
   \includegraphics[width=0.49\columnwidth]{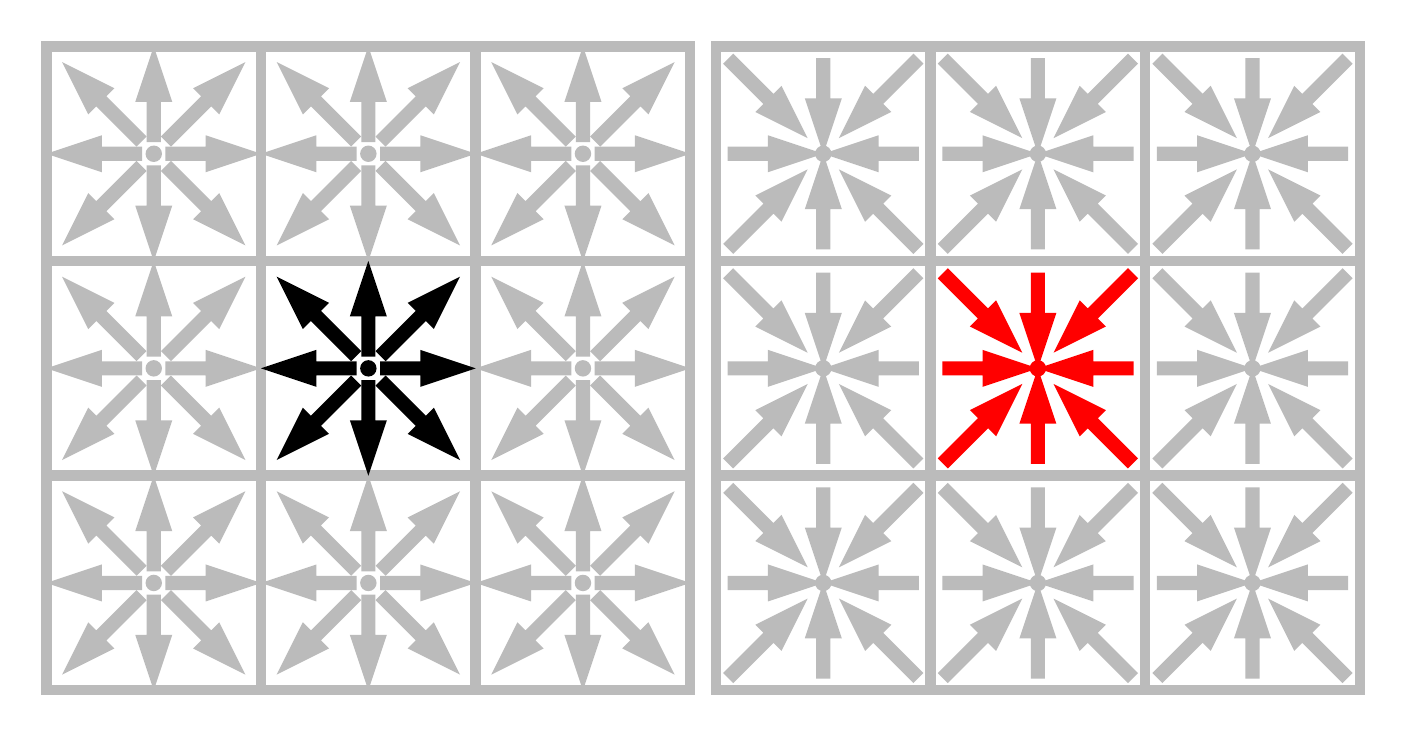}
   \includegraphics[width=0.49\columnwidth]{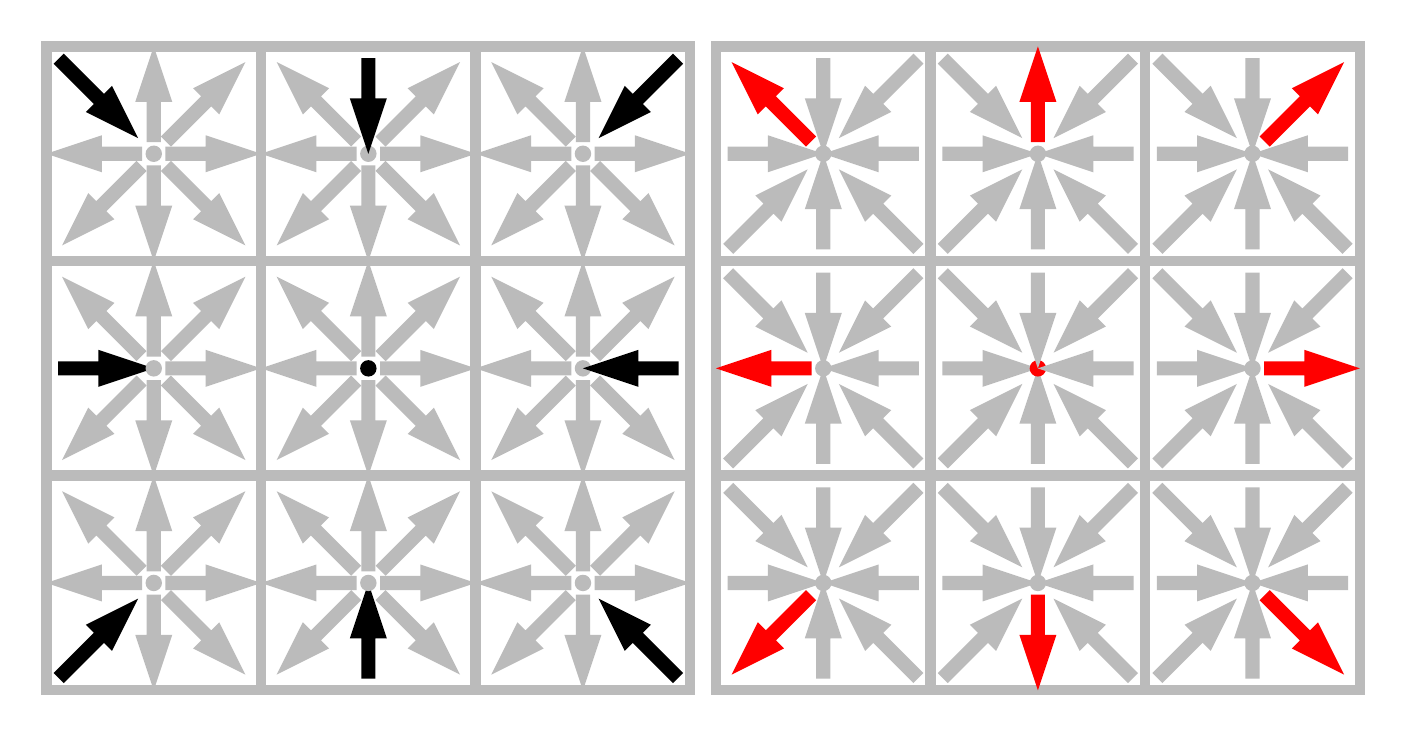} 
 \end{center}
 \caption{AA update pattern. The two leftmost schematics show the even time step consisting of an in-place collision with inversed storage of populations. The odd time step (right) is a fused stream-pull, collide, stream-push step.}
 \label{fig:aa_pattern}
\end{figure}

To be able to process all cells in parallel, while only having a single array for PDF storage, the AA pattern needs two different access patterns for even and odd time steps~(\cref{fig:aa_pattern}).
This also leads to two different kernels that have to be run in an alternating fashion.
The different data layout after even and odd steps may complicate boundary handling and coupling the LBM to other solvers, when traditional implementation techniques are used. 
With our code generation approach this additional complexity can be handled automatically.
The symbolic, high level formulation of method, boundaries, and update scheme is sufficient to e.g. generate boundary handling for even and odd steps automatically.


\begin{figure}[H]
 \begin{center}
   \includegraphics[width=0.49\columnwidth]{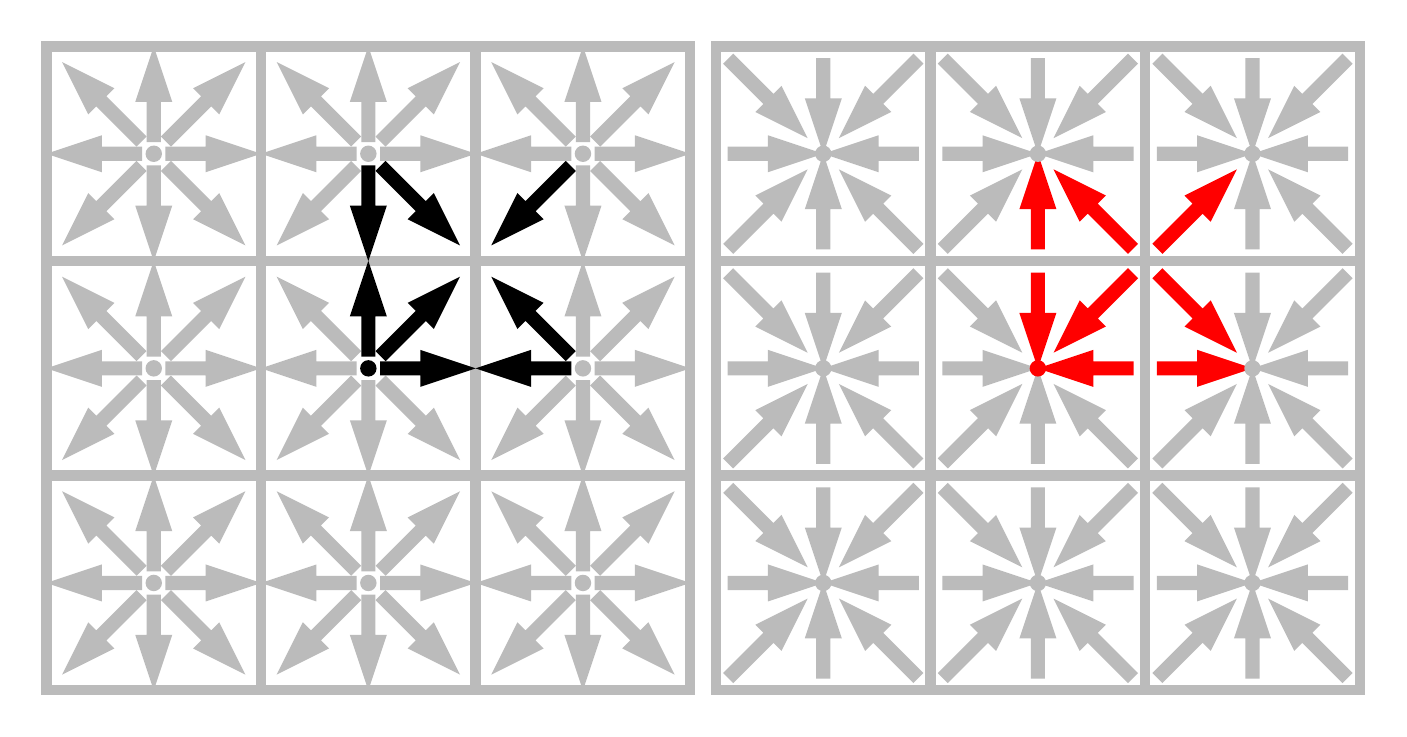}
   \includegraphics[width=0.49\columnwidth]{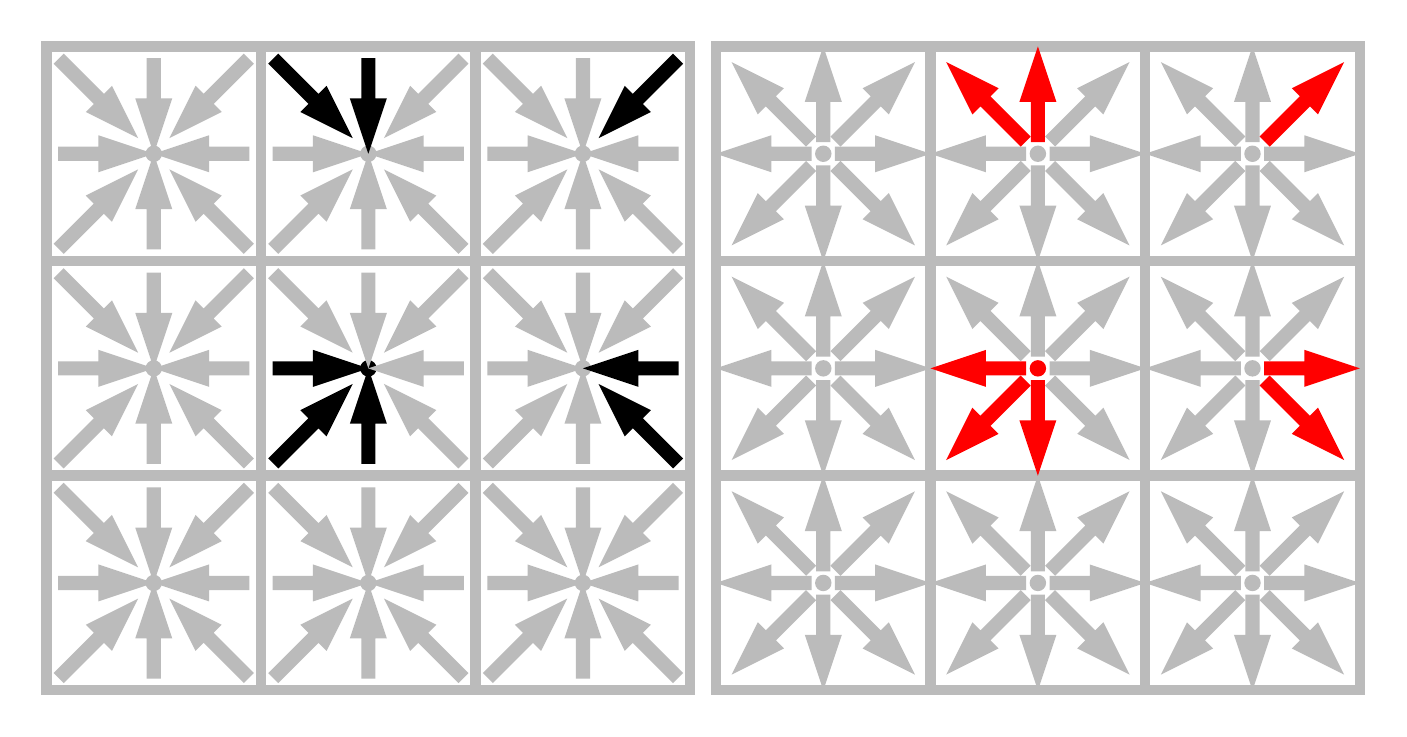} 
 \end{center}
 \caption{Esoteric twist split in even (left) and odd (right) time step kernels.
          Black arrows indicate reads, red arrows writes.}
 \label{fig:esotwist}
\end{figure}

The esoteric twist pattern also requires only a single array.
In contrast to the AA pattern,
it was designed to not require an even and odd time step.
If the populations for different lattice velocities are stored in separate arrays,
a pointer swapping technique can be used for streaming. 
In \lbmpy{}, however, we do not use this technique, in order to keep a common kernel interface with a single population array. 
Instead we also use an even and an odd time step for the EsoTwist pattern as well (\cref{fig:esotwist}).

\subsubsection{Boundary Conditions}

In this section we discuss in more detail how boundary conditions are realized algorithmically. 
One option is to leave the LB kernel unchanged, and run separate boundary handling kernels before. These kernels prepare the population array by writing values that will be streamed in from boundary cells. \lbmpy{} can take symbolic boundary definitions, as shown above, and generate one kernel per boundary. In the simplest case, these boundary kernels operate on a rectangular subdomain, e.g., at the borders of the computational domain. This very simple, but also very common case can thus be handled in the most efficient way possible. 

For more general boundary shapes a flag field is used. 
The flags store a bitmask in every cell that encodes the type of boundary.
The flag field is initialized by the user using image/voxel data or with the help of surface meshes. 
The boundary condition kernel could iterate over the full domain,
masking out cells, but especially if boundary conditions are static, this would be rather inefficient. 
Thus, \lbmpy{} also offers an alternative approach, where instead of iterating over all cells,
a pre-processing step extracts boundary cell coordinates from the flag field and creates a list of indices for
each boundary condition.
This index list contains the spatial coordinate of the boundary cell together with the lattice direction to
the neighboring fluid cell.
Consequently, this list has one entry per boundary link.
For each boundary link, custom boundary data can be stored additionally,
e.g., the wall velocity for a velocity bounce-back boundary.
The flag field then only acts as a convenient way to setup boundaries. 
During the simulation itself, it is not required any more
since all necessary information is stored in the list data structure to accelerate the processing.

So far we have studied
time steps where the boundaries are treated in a separate kernel.
With \lbmpy{}, boundaries can also be compiled directly into the LB compute kernel.
Then a conditional is added to the kernel that determines the cell type
either by using a flag field or by a boolean expression that depends on the spatial coordinates.

Using the information about the data access pattern,
the symbolic formulation is transformed to obtain a concrete boundary assignment.
We point out that it is particularly beneficial to automate the
error prone implementation of boundary conditions for the single-array patterns AA or EsoTwist.

All boundary treatment options discussed above are not new since they have already been
implemented in existing frameworks or applications.
The new contribution here is that code for all these options can be automatically
generated avoiding tedious manual coding and debugging. 
Since all versions can be generated easily, this makes it possible to benchmark all versions
and to choose the fastest version for a specific setup.
Furthermore, there is no trade-off between flexibility and performance any more.
It is now possible to compile a specific  boundary treatment into a kernel to get an
application-specific implementation with the best performance.
Different from a hand-tuned version of the same method, the \lbmpy{} approach keeps the
system maintainable and extensible. 
The separation of concerns is realized on the symbolic abstraction level.


\subsection{Transformations in the Intermediate Representation}

In this section we describe low level optimizations 
executed on the intermediate representation of \pystencils{} 
with the goal to further accelerate the LB compute kernels.
\pystencils{} is designed as a modular package that allows the user
to write custom code transformations specific to the application. 

\subsubsection{Splitting inner loop}
For LB kernels we expect that the memory interface
to be the performance limiting factor, if domains exceed the capacity of the outer level cache. 
The first optimization we discuss here, aims to increase the maximum attainable bandwidth of the kernel by reducing the number of parallel load/store streams from/to memory. 
A standard LB kernel iterates over all cells, loads all $q$ pre-collision values at once,
computes the post-collision values and stores all $q$ of them.
This leads to $q$ parallel load and store streams. 
Reducing the number of parallel streams to memory can increase the obtained bandwidth~\cite{wellein2006towards}.
Therefore we develop an automatic transformation that splits the innermost loop into multiple smaller loops. 
To avoid the re-computation of common subexpressions in every inner loop, 
buffer arrays are introduced.
The first inner loop computes density and velocity and writes them to the buffer arrays. 
The following loops then handle only two lattice direction updates and have only two parallel load and store streams.
Algorithm~\ref{alg:splitlbm} shows the state after the transformation in pseudo-code.
It assumes a simple two-field storage pattern with source and destination array.
\begin{algorithm}
\begin{algorithmic} 
 \For{all slices $y, z$}
    \State ρ\_arr $\leftarrow $ array[x-size]
    \State u\_arr $\leftarrow $ array[x-size]
    \For{line $x$}
         \State $f \leftarrow \mbox{src}[x, y, z]$
         \State ρ\_arr[x] $\leftarrow ρ(\mathbf{f})$
         \State u\_arr[x] $\leftarrow \mathbf{u}(\mathbf{f})$ 
         \State dst[x, y, z, center] $\leftarrow Ω(f, \mbox{ρ\_arr[x]}, \mbox{u\_arr[x]})$ 
    \EndFor
    \State
    \For{line $x$}
        \State $f \leftarrow \mbox{src}[x, y, z]$
        \State dst[x, y, z, east] $\leftarrow Ω(f, \mbox{ρ\_arr[x]}, \mbox{u\_arr[x]})$ 
        \State dst[x, y, z, west] $\leftarrow Ω(f, \mbox{ρ\_arr[x]}, \mbox{u\_arr[x]})$         
    \EndFor
    \For{line $x$}
        \State $f \leftarrow \mbox{src}[x, y, z]$
        \State dst[x, y, z, north west] $\leftarrow Ω(f, \mbox{ρ\_arr[x]}, \mbox{u\_arr[x]})$ 
        \State dst[x, y, z, south east] $\leftarrow Ω(f, \mbox{ρ\_arr[x]}, \mbox{u\_arr[x]})$         
    \EndFor    
    \State ... {\em (more loops for remaining directions)}
 \EndFor
\end{algorithmic}
\caption{Stream-collide kernel with split inner loops}
\label{alg:splitlbm}
\end{algorithm}
There are different options on how to exactly 
perform this transformation. 
One free parameter is the number of directions that are updated in the inner loops. 
In the example, we update two opposing directions at once, since these updates share many common subexpressions.
One could also create a separate inner loop for each direction, or group more than two directions together.
This transformation is also parametrized by the common subexpressions that are pre-computed in temporary arrays.
\lbmpy{} can introduce additional temporary arrays for other subexpressions besides density and velocity as well.
A heuristic is used to determine subexpressions
 that are compute intensive enough to justify introducing a temporary array for them.
It is important that all temporary arrays fit into the inner level cache so that they do
not generate additional pressure on the memory interface.
In the simple example, as shown in algorihm~\ref{alg:splitlbm},
the temporary arrays grow with the domain size in $x$-direction.
To make this optimization work for arbitrary domain sizes, the inner loop is blocked before splitting it up.
The chunk size can be selected such that the arrays fit into L1 cache.

\subsubsection{OpenMP and SIMD vectorization}
All LB kernels are designed in a way that cells can be updated in parallel. 
{\em pystencils} uses the fact that loop iterations are independent to automatically parallelize the kernel with OpenMP.
By default the outer loop is parallelized using a static scheduling strategy.
If the domain size is known at compile time, and the outer dimension is very small,
{\em pystencils} uses OpenMP \texttt{collapse} to increase the number of parallel interations. 

Knowing that iterations are independent, allows {\em pystencils} to vectorize the code. 
To have full control over the vectorization process, we do not rely on compiler auto vectorization 
or pragma-based approaches but generate C code with SIMD intrinsics.
{\em pystencils} currently supports SSE, AVX, AVX2 and AVX512 vector instruction sets. 
If the data layout and alignment of the population array is known at compile time,
we generate aligned load/store instructions where possible. 

\subsubsection{Non-temporal stores}
The intrinsics-based vectorization allows us to explicitly use
{\em non-temporal} (NT) stores, also called {\em streaming stores}, in kernels that use two population arrays.
This optimization reduces the total amount of data that has to be transferred from/to memory.
By default, modern CPUs have a ``write-allocate'' or ``read for ownership'' cache policy~\cite{Wittmann2013a}. 
This means that a store operation causes the respective cache line to be read
into cache and thus generates twice the memory traffic that is actually required.
This actually causes $3q$ values to be transferred over the memory interface per lattice cell.
The custom vectorization allows us to change the store instructions from default to streaming stores
that bypass the cache. 
Then only $2q$ values have to be loaded and stored per cell.
So this optimization can increase the performance of two-array LB kernels by a factor of 1.5,
assuming they are memory-bound and the PDF array does not fit into the outer level cache.

\subsection{Framework Integration} 

The intermediate representation of the compute- and boundary kernels is finally transformed by 
a backend to either C, CUDA or OpenCL code.
For each kernel a C function with a well-defined interface is generated. 
Arrays are passed in as raw pointer, together with shape and stride information that define the memory layout of the arrays. Symbolic quantities that have not been replaced during the code generation process automatically become parameters to the generated C function, e.g., values for relaxation rates or constant external forces. 

This simple interface was chosen, such that the generated kernels can be called from a variety of different languages and can be easily integrated into existing frameworks.
In this section we describe different ways of utilizing {\em lbmpy}. 
The first option allows the user to completely work in a Python environment,
preferably an interactive Jupyter notebook for a convenient display of symbolic expressions.
There, the user derives the LBM symbolically and passes the method definition to {\em lbmpy}.
After automatic simplification and optimization the generated C/CUDA/OpenCL code is
automatically compiled and dynamically loaded as a Python module.
The compilation process is fully transparent to the user. 
The optimized, shared-memory parallel kernel can then be directly called from Python.
Data is stored in {\em numpy} for CPU simulation or in {\em gpuarray}'s from the {\em pycuda} package for GPU simulations.
In this mode \lbmpy{} offers a flexible and fast prototyping environment for LB methods, where simulations can be run on a single node or a single GPU. 

For distributed memory parallelization we use the \walberla{} framework \cite{feichtinger2011walberla},
a multiphysics software system that is optimized for massively parallel simulations with stencil codes.
The distributed memory parallelization uses a block-structured domain partitioning based on a forest of octrees
and is characterized by excellent scalabilty since it uses no central nor globally shared data structures
so that no global communication is necessary \cite{schornbaum2016massively}.
This fully parallel data structure enables adaptive grid refinement and dynamic load balancing between
MPI processes~\cite{schornbaum2018extreme, bauer2020walberla}. 
\walberla{} has Python bindings~\cite{Bauer2016python} that allow for simple distributed simulations
with \lbmpy{} generated kernels directly from Python on a uniform grid. 
For advanced use cases, e.g., those that require grid refinement, 
the user has to switch to C++ as the driving language. 
Integrations of \lbmpy{} into the CMake build system of \walberla{} control the generation of LB compute kernels, boundary kernels and packing/unpacking kernels for distributed memory MPI communication. 
The reason why we also generate communication kernels are the single-field AA and EsoTwist storage patterns.
Manually determining what values have to be sent to neighboring processes is tedious and error-prone in these cases.
Since the compute kernels are available in symbolic form, we can extract that information and generate
the necessary communication routines automatically.



\section{Performance Results}
\label{sec:performance}

In this section we present benchmark results using the automatically generated LB kernels
and compare them to manual implementations with different optimization level. 

\subsection{Single Node Benchmark}

\subsubsection{Hardware}
We first investigate the single-node performance on two test systems.
We scale all kernels on one socket of two Intel Xeon processors with different microarchitecture.
The first system is an Intel Xeon E5-2695v3 Haswell system with 14 physical cores per socket. 
For benchmarking, we deactivate the turbo mode of this processor and set the frequency to a fixed value of
2.3 GHz using the likwid tool suite~\cite{treibig2010likwid}.
The second system is an Intel Xeon Gold 6148 CPU Skylake that has 20 cores per socket with a fixed frequency of 2.4 GHz. The Sub-NUMA clustering features of both systems are switched off to have one NUMA domain per socket.
We use transparent huge pages and disable automatic NUMA balancing in the Linux kernel. 
For all benchmarks we use a domain size of $300 \times 100 \times 100$ that is too large
to fit in the outer level cache of any of the tested systems.

To find an upper bound for the possible performance,
assuming kernels are memory-bound, we use bandwidth measurements for both systems from~\cite{Wittmann2018}. 
\Cref{tbl:systembw}~shows the measured \texttt{copy} bandwidth for both machines
where write-allocates have already been taken into account. 
Additionally, we use bandwidth measurements of scenarios that closely mimic the memory
access behaviour of the D3Q19 LB kernels. 
Kernels with two-array population storage are compared to a stream benchmark with
19 parallel streams and non-temporal stores,
labeled \texttt{copy-19-nt-sl}.
Kernels with a single-array update pattern are compared to a benchmark that updates
19 values in place called \texttt{update-19}. 
The upper bound for the kernels, as measured in million lattice updates per second (MLUP/s),
the bandwidth is divided by the number of bytes that have to be transferred per lattice cell.
We assume double precision for all kernels.
Thus each cell update requires $2\cdot q \cdot 8$ bytes per cell.

\begin{table}
\centering
 \begin{tabular}{l|cc}
    processor                 & Xeon E5-2695v3    &  Xeon Gold 6148       \\
    \hline
    micro architecure         & Haswell           &  Skylake              \\
    cores per socket          & 14                &  20                   \\
    frequency                 & 2.3 GHz           &  2.4 GHz              \\
    \hline
    {\em Measured Bandwidths} &                   &                       \\
    \texttt{copy}             & 52.0 GB/s         &  102.8 GB/s           \\
    \texttt{copy-19-nt-sl}    & 47.1 GB/s         &  92.4 GB/s            \\
    \texttt{update-19}        & 44.0 GB/s         &  93.6 GB/s            \\
 \end{tabular}
 \caption{Test system specification with measured bandwidths from~\cite{Wittmann2018}. 
   \texttt{copy} uses one load and store stream, write-allocate is already taken into account. \texttt{copy-19-nt-sl} uses 19 load and store streams and non-temporal stores. \texttt{update-19} updates 19 values in-place.}
 \label{tbl:systembw}
\end{table}

The benchmark codes are compiled with Intel compiler 19.0.2 if not specified otherwise. 
Where explicitly noted, the GCC in version 7.4.0 is used. 
For both compilers we set the optimization flag \texttt{-O3}, enable AVX512 on Skylake and AVX2 on Haswell, and switch on fast math flags that allow the compiler to reorder floating point operations.
For the intel compiler these are \texttt{-fp-model fast=2 -no-prec-sqrt -no-prec-div} and for GCC \texttt{-ffast-math}.
All kernels are parallelized with OpenMP. 

\subsubsection{Two-array kernels and comparison to manual implementations}
 
As discussed before, there is a trade-off between code quality and performance when
developing LB kernels manually in C/C++. 
Code is optimized by specializing it for a particular scenario.
To illustrate this trade-off we compare {\em lbmpy} generated kernels with two manual implementations.
The code of the manual implementations can be accessed at~\cite{manualImplGithub}.  
For this test we restrict ourselves to a SRT collision operator.
The first manually implemented code is written in a stencil-agnostic way,
where lattice velocities and stencil weights are abstracted away through template meta-programming.
Theoretically, the compiler should be able to resolve these indirections fully at compile-time.
Relaxation rates are also passed in via a templated functor, to enable a flexible integration of turbulence models.
While the main aim of this kernel is to be as generic as possible it is still restricted to a single collision operator and a single two-array population storage pattern.
But it is easily readable and extensible. 

\begin{figure*}[h]
 \centering
 \includegraphics[width=0.9\textwidth]{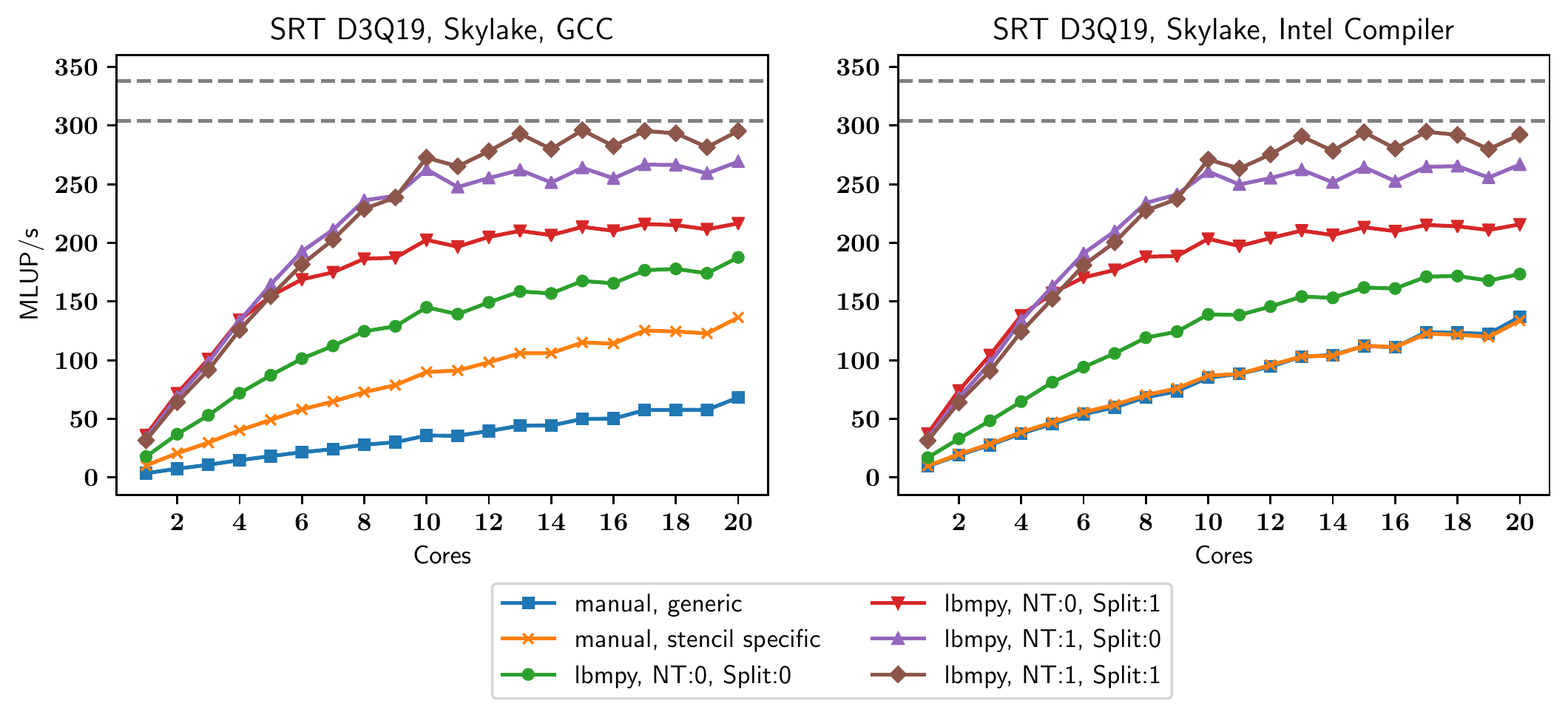}  
 \caption{Comparison of kernels with different optimization level on Skylake using a BGK method with two-array population storage on a $300 \times 100 \times 100$ domain. Horizontal lines indicate the roofline estimate using the measured \texttt{copy-19-nt-sl} (lower) and \texttt{copy} bandwidth (higher).} 
 \label{fig:two_field_skylake}
\end{figure*}

The second manual implementation is written specifically for a D3Q19 stencil.
All loops over lattice directions are manually unrolled,
expressions are simplified by leaving out multiplications with zero lattice direction components,
and common subexpressions are eliminated.
These steps lead to code duplication and decreased readability, but may lead to better performance. 
These optimization steps should not be necessary since the compiler should be able to do them automatically.
However, the unrolled stencil-specific version is the basis of further optimization like loop splitting.
\Cref{fig:two_field_skylake} shows benchmark results on the Skylake system for both manually implemented kernels using GCC and the Intel compiler.
We can see that indeed the Intel compiler (right) was able to resolve the compile-time abstractions, such that the generic version is as fast as the stencil-specific one.
However, GCC cannot optimize the generic code automatically, only obtaining about half the performance. 
The manual implementations scale perfectly, but are far from utilizing the available bandwidth on the system. 
The generated kernel without loop splitting and non-temporal stores already performs
better than the manual implementations.
This kernel is explicitly vectorized with AVX512 SIMD intrinsics and uses pointer arithmetic to access the population arrays, whereas the manual implementations use getter/setter methods of an array class. 
Splitting the inner loops lets the kernel saturate at about 200 MLUP/s.
Due to the write-allocate strategy in total 1.5 times more data is moved across the memory interface than necessary.
As can be seen in the plot, the performance of this kernel is consequently also about a factor of 1.5
worse than the best kernel with NT stores.
Activating NT-stores results in the expected performance of about 300 MLUP/s on this system,
very close to the maximal 304 MLUP/s predicted by the roofline estimate obtained with the \texttt{copy-19-nt-sl} bandwidth.
So the loop splitting and non-temporal stores optimizations are indeed necessary to obtain best possible performance on this system.
All manual implementations, where these optimizations have been applied,
are lengthy and hard to read, due to the unrolled loops and the usage of SIMD intrinsics.
These hand-optimized codes are not only difficult and time-consuming to develop,
but also their maintainability and flexibility have been sacrificed for performance.
With code generation it is possible to resolve these conflicting goals.
\Cref{fig:two_field_skylake} also shows, that the generated code performs consistently across different compilers,
since all abstractions are already transformed to perform with best possible efficiency
by the code generation system, leaving only standard optimizations to the back end compiler.

\begin{figure}
 \centering
  \includegraphics[width=0.9\columnwidth]{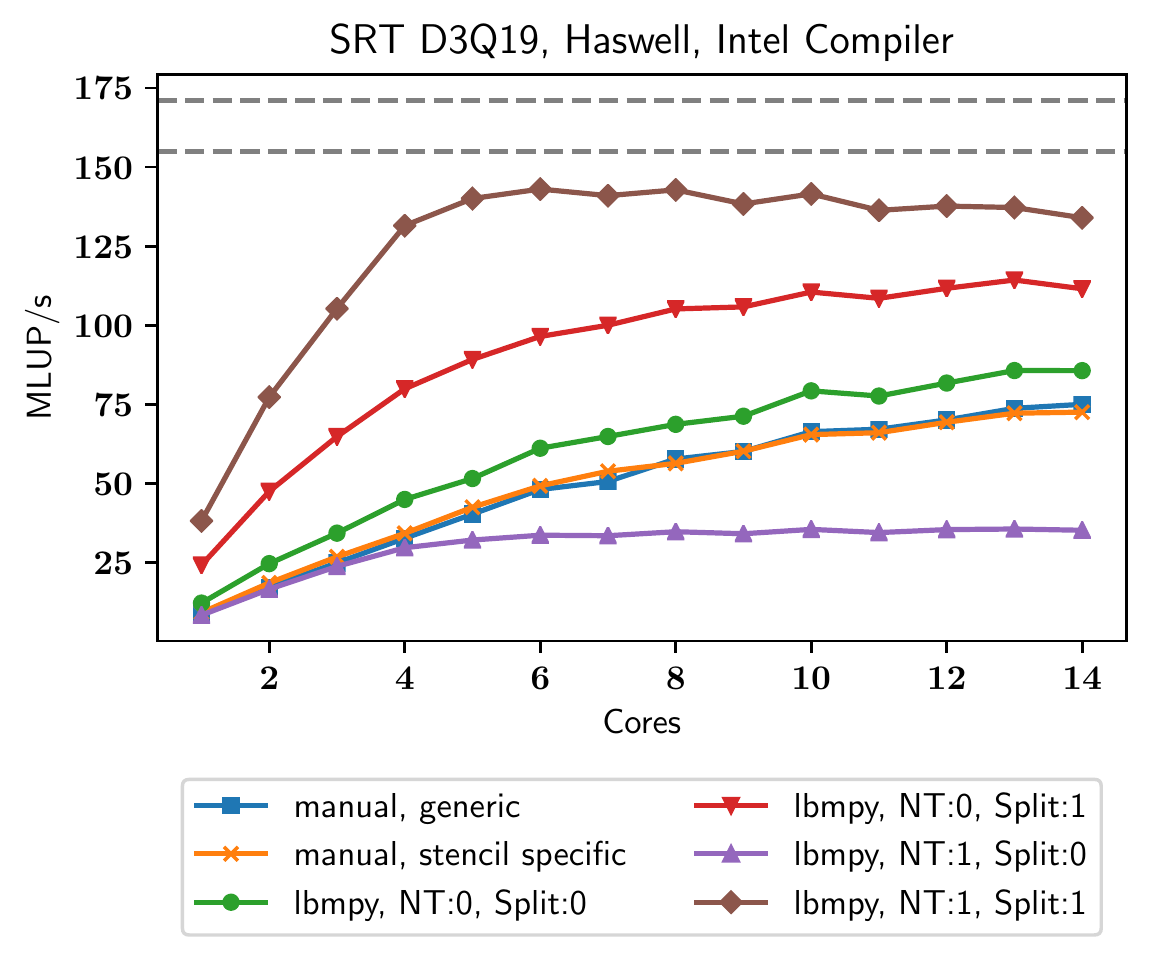}
  \caption{Two-field BGK D3Q19 kernels on Haswell. Configuration and roofline indicators as in~\cref{fig:two_field_skylake}. }
  \label{fig:two_field_haswell}
\end{figure}

\Cref{fig:two_field_haswell} shows the corresponding results for the Haswell system.
Overall the behavior of this older system is similar to Skylake, with the exception that the system is apparently not able to handle $19$ parallel non-temporal store streams, as the version with NT-stores without loop splitting performs very poorly.

\subsubsection{Kernels with AA pattern and boundary handling}
\begin{figure}
   \centering
    \includegraphics[width=\columnwidth]{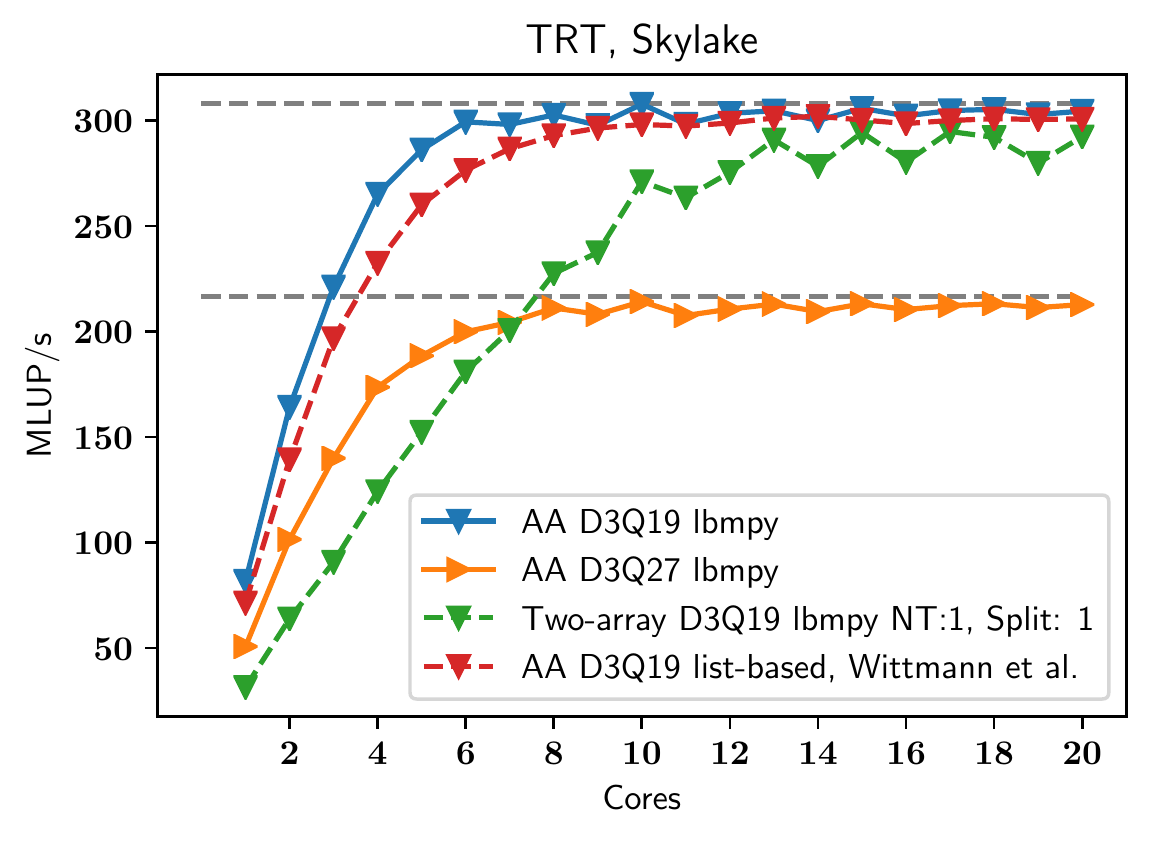}
    \caption{AA pattern, TRT on Skylake. Roofline estimate uses \texttt{update-19} bandwidth.}
    \label{fig:aa_skylake_kernel_only} 
\end{figure}

Next, we show performance results for single-array kernels that use the AA update pattern.
We use the TRT collision operator for these benchmarks.
SRT and TRT kernels have very similar performance characteristics, because they have about the same number of FLOPs. 
In \cref{fig:aa_skylake_kernel_only} we compare the best two-field version with split loops and non-temporal stores to the corresponding AA kernel.
The two-array kernel saturates at about 13 cores, the AA version achieves the highest performance already with 6 cores.
The additional development effort that is required for the AA pattern pays off not only in half the memory consumption but also in single core performance.

For the AA kernels the NT-store optimization is not applicable, since all values are updated in-place. 
The inner loop splitting, however, may be beneficial. 
Since there are two different kernels for even and odd time steps, there are in total four options, where the splitting transformation has been applied to none, only one, or both kernels.
We find, that loop splitting does not help in this case.
All four options yield almost equal performance results.
Thus, \cref{fig:aa_skylake_kernel_only} only shows the version where neither of the two kernels has been split.

We can also see, that the roofline limit based on the measured \texttt{copy-19} bandwidth is a very good model for the performance on the full socket. 
The D3Q27 version saturates at very close to the expected value, that is by a factor of $19/27$ lower than that of the D3Q19 stencil.

\Cref{fig:aa_skylake_kernel_only} also shows performance results of the TRT LB benchmark kernel by Wittmann et al.~\cite{Wittmann2018}. 
From this benchmark we use the fastest kernel~\texttt{list-aa-pv-soa} on a channel geometry. 
It also uses the AA pattern in a SoA layout. 
In contrast to the \lbmpy{} kernels, it operates on a sparse list data structure, such that only populations
in fluid cells have to be stored.
Also, the benchmark kernels have boundary handling built in,
while the \lbmpy{} results in \cref{fig:aa_skylake_kernel_only} show the performance of the compute kernel only.

\begin{figure}
   \centering
    \includegraphics[width=\columnwidth]{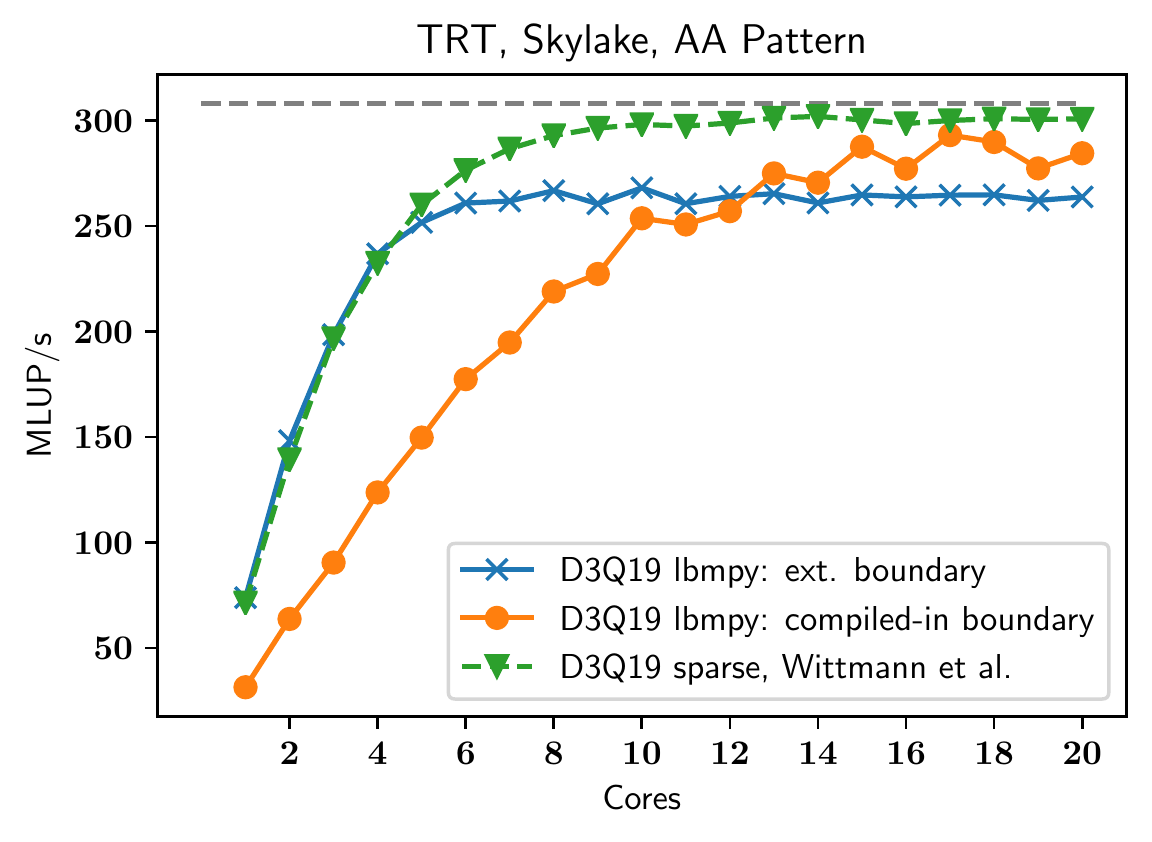}
    \caption{TRT collision operator on Skylake using AA pattern with different boundary options.}
    \label{fig:aa_skylake_with_boundary}
\end{figure}

The boundary handling performance of \lbmpy{} is investigated in \cref{fig:aa_skylake_with_boundary}. 
It shows the \lbmpy{} kernels where two different boundary handling approaches are used for a channel scenario, where non-periodic boundaries are set on all sides.
The simplest option is to generate separate, external kernels that handle boundaries.
Since cache lines containing population at the border must be loaded twice during a time step,
the final performance obtained at the full socket is decreased by about 15\%.
The second approach introduces conditionals in the compute kernel.
With this approach we can obtain about the same performance on the full socket as the pure compute kernel, but the performance on a few cores is much lower. 
The intrinsics-based SIMD vectorization in {\em pystencils} cannot handle the conditionals in an optimal way yet.
This limitation is expected to be remedied in future work. 

\subsubsection{Advanced collision operators}
\begin{figure*}
   \centering
    \includegraphics[width=\textwidth]{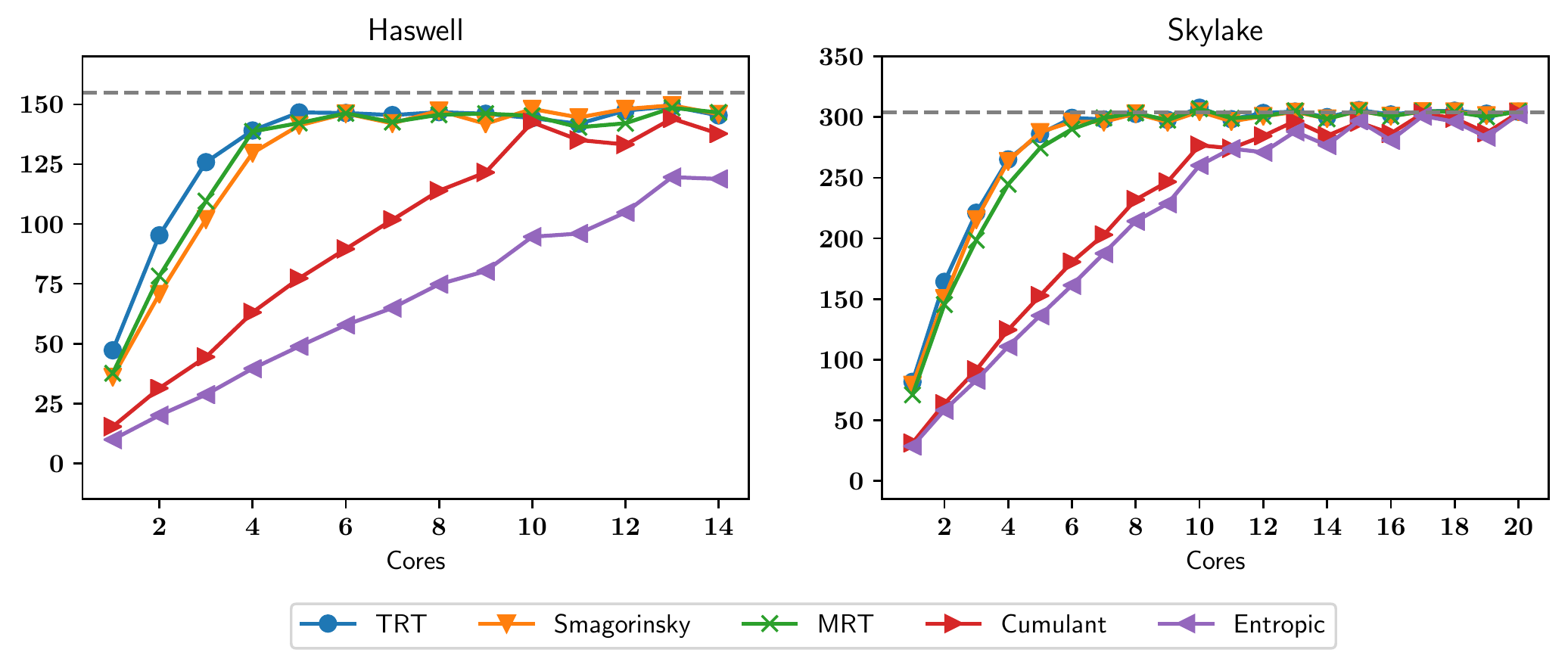}
    \caption{Comparison of different LB collision operators. 
             All kernels use the AA pattern and a D3Q19 stencil.}
    \label{fig:aa_method_comparison} 
\end{figure*}

The goal of code generation in \lbmpy{}  is not to make a single collision operator fast,
but provide a framework that is can obtain good performance for a wide range of different LBMs.
\Cref{fig:aa_method_comparison} shows results for different D3Q19 LB schemes on the test systems using the AA update pattern. 
The TRT results, we have seen above are included for reference again. 
A slightly more complex scheme is the BGK operator with included Smagorinsky turbulence model. 
In this kernel, the relaxation rate is determined on a cell-by-cell basis.
The computation of the adapted rate is done on the fly inside the kernel, to not introduce additional memory accesses. Schemes with variable relaxation rates are oftentimes implemented in a way where the rates are
computed in a separate kernel and stored into an additional array,
for flexibility reasons.
This is not necessary in \lbmpy{}, so that the Smagorinsky kernel obtains identical performance as the TRT kernel on the full socket.
On Skylake also the performance on small core counts is almost identical to the TRT kernel,
whereas on Haswell the additional computational complexity,
like e.g., the two sqrt operators per cell, lead to lower single core performance compared to TRT. 

Next, we investigate performance characteristics of an MRT kernel with weighted orthogonal moments.
It has four relaxation rates, two for controlling shear and bulk viscosity separately, one for third, and one for forth order moments. 
All relaxation rates remain symbolic at compile time and become run time parameters. 
\lbmpy{} is capable to optimize this model so that it almost runs as fast as the TRT model on both test architectures.
This is true also for other MRT models that the system can generate, e.g. weighted/unweighted moment orthogonalization or compressible/incompressible equilibria.

\Cref{fig:aa_method_comparison} also contains measurements for a D3Q19 cumulant method. 
The non-linear transformation to cumulant-space makes this collision operator more compute intensive than MRT methods. Nonetheless, \lbmpy{} can optimize the cumulant kernel such that it 
saturates the available memory bandwidth on both systems. 
Additionally, we try an entropic method of KBC type.
Shear and bulk viscosity is kept fix,  the relaxation rate for higher order moments is chosen adaptively to maximize entropy. 
This method is too compute intensive to be memory-bound on Haswell, but on a full Skylake socket it achieves 
performance similar to the simpler methods. 

Summarizing our findings, after careful optimization there is no performance penality
using complex LB collision operators.
On modern CPU architectures all LBM implementaions are memory bound when they are properly optimized.
This optimization, however, is only achievable by tedious manual coding by experts
or by using automatic code generation with tools like \lbmpy{}.



\subsection{Scaling Benchmark}

Integrating the generated \lbmpy{} kernels into the \walberla{}
framework allows us to run large scale simulations on distributed memory systems. 
We use the MPI communication capabilities of \walberla{} together with generated 
serialization/deserialization kernels to run a large parallel simulations. 
In contrast to previous work~\cite{walberlaSC13,BAUER2020}, where scaling
results for manual implementations of TRT two-field kernels have been shown, 
we demonstrate the performance of a more complex MRT kernel with AA pattern here. 
As we have shown above, this kernel runs as fast as a SRT or TRT collision operator
on the full node when properly optimized. 

As test system the SuperMUC-NG supercomputer in Munich is used. 
It consists out of Intel Xeon Platinum 8174 processors with Skylake architecture. 
Each node has two sockets with 24 physical cores each.
We run a weak scaling setup up to the 3072 compute nodes that we have access to. 
This is about half of the full machine size.

\begin{figure}
   \centering
    \includegraphics[width=\columnwidth]{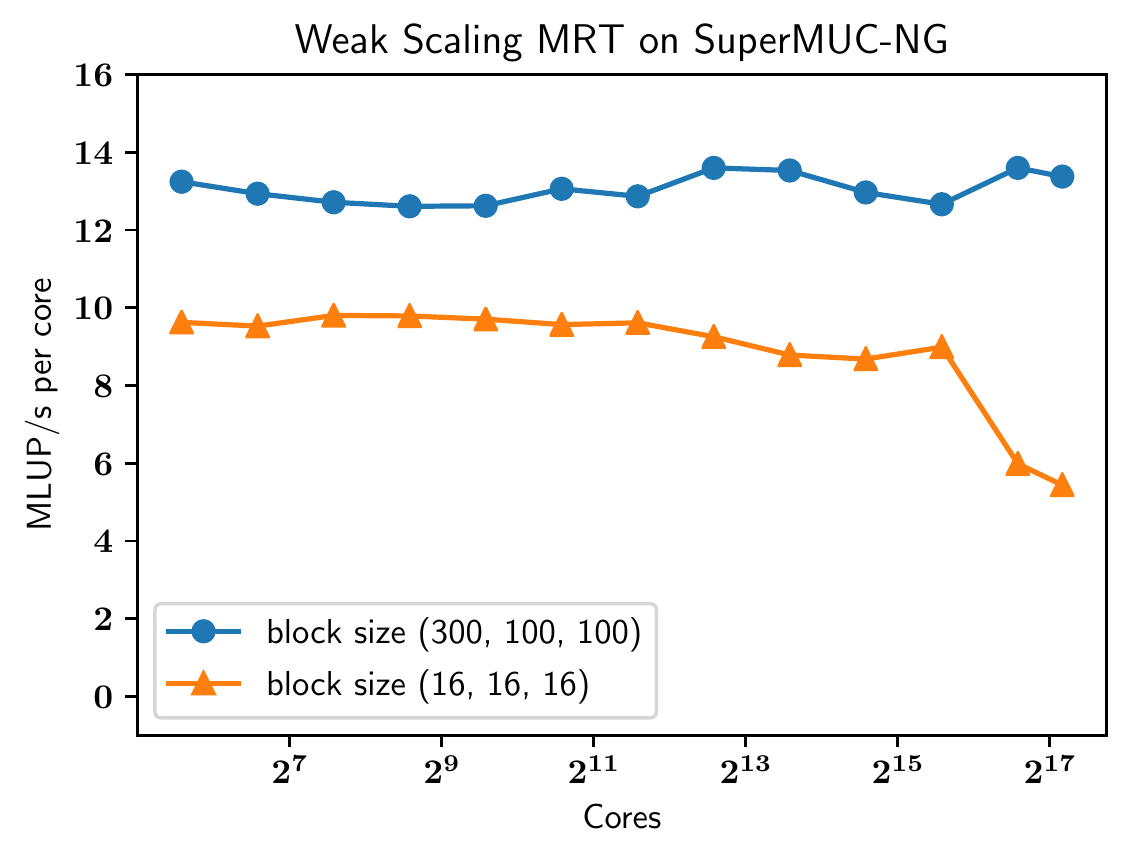}
    \caption{Weighted orthogonal MRT method with AA update pattern scaled on SuperMUC-NG up to 147,456 cores (3072 nodes) using a channel geometry.}  
    \label{fig:supermuc-ng-scaling}
\end{figure} 

\Cref{fig:supermuc-ng-scaling} shows the scaling results of a channel geometry.
The domain is partitioned into equally sized blocks and each block is assigned to a physical core. 
The machine is best utilized when choosing large block sizes, since the communication overhead is kept small in this case.
For a block size of $(300, 100, 100)$, we observe perfect scalability up to all 147,456 cores used. 
In this configuration we obtain about 1972 GLUP/s on half of SuperMUC-NG. 
Besides this weak scaling scenario demonstrating maximal GLUP/s rates, we
may alternatively want to maximize the number of time steps per second. 
To illustrate the capabilities of \lbmpy{} with \walberla{} we also study a scaling scenario with much
smaller block size.
\cref{fig:supermuc-ng-scaling} shows the scaling behavior for a block size of $16^3$.
Here we achieve still  good scalability up to 1024 nodes,
then the performance in GLUP/s drops to approximately 40\% of the performance that was observed for the 
large block size. 
Note however, that in this configuration we can execute 1327 time steps per second. 
Note also that on the 3072 nodes on SuperMUC-NG this is still for an LBM grid consisting of $6 \times 10^8$ LBM cells.

\section{Conclusion and Outlook}
\label{sec:conclusion}

In this article, we presented a programming system named \lbmpy{} 
that supports the flexible creation of highly optimized parallel LBMs.
The scope of \lbmpy{} are moment-based MRT methods plus cumulant and entropically stabilized collision operators.
All methods can be created with locally varying relaxation parameters
so that various turbulence models can be realized or also models for non-Newtonian fluids.
\lbmpy{} automatically optimizes the compute kernels with
domain-specific transformations and it can produce codes that 
employ memory-efficient single-array population storage.
Even for complex LBM models, the kernels generated by \lbmpy{} can reach the same performance as
manually optimized state-of-the-art TRT implementations of~\cite{Wittmann2018}.
After automatic optimization, all methods are memory-bound on a recent Skylake system.
Thus, also advanced collision operators can be used without performance penalty, provided
that large domain sizes are used and memory bandwidth is the bottleneck.
Through integration into the HPC framework \walberla{}, 
the \lbmpy{}-generated methods can be executed 
on large scale distributed-memory systems with excellent scalability.  

The \lbmpy{}/\walberla{} programming system supports the computational science workflow.
It automizes the tedious and error prone development steps.
The support of \lbmpy{} is {\em vertically integrated}:
it starts with the development of advanced kinetic schemes,
it assists the development of scalable parallel codes, and it 
includes advanced hardware-specific code optimizations.
Note that the derivation of modern LB schemes will usually require 
expertise in mathematics and physics by specialists in LBMs,
while the nodel-level kernel optimization for modern CPU microarchitecture will
require a detailed understanding of CPU microarchitecture.
In this sense \lbmpy{}/\walberla{} is an excercise in interdisciplinary co-design
to create advanced simulation software for future extreme schale computing.
The code-generation paradigm permits
a higher level of abstraction than it can be realized
by conventional software engineering methods.
These new, application-specific methods of abstractions can only be realized with
automatic code generation.
They help to resolve the fundamental conflict between flexibility of software
and the need for hardware specific optimizations.
The approach taken in \lbmpy{} offers a road to performance portability and thus
to improve the sustainability of scientific software.

\section{Acknowledgements}

The authors are grateful for funding received through the project HPC2SE (01ICH16003D) from the Bundesministerium für Bildung und Forschung (BMBF).
Additionally we are grateful to the Regionales Rechenzentrum Erlangen, the Leibniz-Rechenzentrum in Garching and the Swiss National Supercomputing Centre in Lugano for providing compute time on their HPC systems.

%
%
%
%
%
%

\bibliographystyle{abbrvdin}
\bibliography{library,manual_library}

\end{document}

%% file: listings/output/00-trt-example.py.tex
\begin{Verbatim}[commandchars=\\\{\},fontsize=\footnotesize]
\PY{n}{d2q9} \PY{o}{=} \PY{n}{get\PYZus{}stencil}\PY{p}{(}\PY{l+s+s2}{\PYZdq{}}\PY{l+s+s2}{D2Q9}\PY{l+s+s2}{\PYZdq{}}\PY{p}{)}
\PY{n}{moments} \PY{o}{=} \PY{n}{independent\PYZus{}raw\PYZus{}moments}\PY{p}{(}\PY{n}{d2q9}\PY{p}{)}
\PY{err}{ω}\PY{n}{\PYZus{}e}\PY{p}{,} \PY{err}{ω}\PY{n}{\PYZus{}o} \PY{o}{=} \PY{n}{symbols}\PY{p}{(}\PY{l+s+s2}{\PYZdq{}}\PY{l+s+s2}{ω\PYZus{}e, ω\PYZus{}o}\PY{l+s+s2}{\PYZdq{}}\PY{p}{)}
\PY{err}{ω}\PY{n}{s} \PY{o}{=} \PY{p}{[}\PY{err}{ω}\PY{n}{\PYZus{}e} \PY{k}{if} \PY{n}{is\PYZus{}even\PYZus{}moment}\PY{p}{(}\PY{n}{m}\PY{p}{)} \PY{k}{else} \PY{err}{ω}\PY{n}{\PYZus{}o}
      \PY{k}{for} \PY{n}{m} \PY{o+ow}{in} \PY{n}{moments}\PY{p}{]}
\PY{n}{m\PYZus{}eq} \PY{o}{=} \PY{n}{maxwellian\PYZus{}moments}\PY{p}{(}\PY{n}{moments}\PY{p}{,} \PY{n}{dim}\PY{o}{=}\PY{l+m+mi}{2}\PY{p}{,} 
                          \PY{n}{c\PYZus{}s}\PY{o}{=}\PY{l+m+mi}{1}\PY{o}{/}\PY{n}{sqrt}\PY{p}{(}\PY{l+m+mi}{3}\PY{p}{)}\PY{p}{)}
\PY{n}{trt} \PY{o}{=} \PY{n}{create\PYZus{}method}\PY{p}{(}\PY{n}{d2q9}\PY{p}{,} \PY{n}{moments}\PY{p}{,} \PY{err}{ω}\PY{n}{s}\PY{p}{,} \PY{n}{m\PYZus{}eq}\PY{p}{)}
\end{Verbatim}

%% file: listings/output/01-mrt-example.py.tex
\begin{Verbatim}[commandchars=\\\{\},fontsize=\footnotesize]
\PY{n}{moments} \PY{o}{=} \PY{n}{independent\PYZus{}raw\PYZus{}moments}\PY{p}{(}\PY{n}{d2q9}\PY{p}{)}
\PY{n}{moments} \PY{o}{=} \PY{n}{split\PYZus{}shear\PYZus{}bulk\PYZus{}moments}\PY{p}{(}\PY{n}{moments}\PY{p}{)}
\PY{n}{moments} \PY{o}{=} \PY{n}{gram\PYZus{}schmidt}\PY{p}{(}\PY{n}{moments}\PY{p}{,} \PY{n}{d2q9}\PY{p}{,} 
                       \PY{n}{weights}\PY{o}{=}\PY{n}{get\PYZus{}weights}\PY{p}{(}\PY{n}{d2q9}\PY{p}{)}\PY{p}{)}
\PY{err}{ω} \PY{o}{=} \PY{n}{symbols}\PY{p}{(}\PY{l+s+s2}{\PYZdq{}}\PY{l+s+s2}{ω\PYZus{}:4}\PY{l+s+s2}{\PYZdq{}}\PY{p}{)}
\PY{err}{ω}\PY{n}{s} \PY{o}{=} \PY{p}{[}\PY{l+m+mi}{0}    \PY{k}{if} \PY{n}{get\PYZus{}order}\PY{p}{(}\PY{n}{m}\PY{p}{)} \PY{o}{\PYZlt{}} \PY{l+m+mi}{2}   \PY{k}{else}
      \PY{err}{ω}\PY{p}{[}\PY{l+m+mi}{0}\PY{p}{]} \PY{k}{if} \PY{n}{is\PYZus{}shear\PYZus{}moment}\PY{p}{(}\PY{n}{m}\PY{p}{)} \PY{k}{else}
      \PY{err}{ω}\PY{p}{[}\PY{l+m+mi}{1}\PY{p}{]} \PY{k}{if} \PY{n}{is\PYZus{}bulk\PYZus{}moment}\PY{p}{(}\PY{n}{m}\PY{p}{)}  \PY{k}{else}
      \PY{err}{ω}\PY{p}{[}\PY{n}{get\PYZus{}order}\PY{p}{(}\PY{n}{m}\PY{p}{)}\PY{o}{\PYZhy{}}\PY{l+m+mi}{1}\PY{p}{]} 
      \PY{k}{for} \PY{n}{m} \PY{o+ow}{in} \PY{n}{moments}\PY{p}{]}
\PY{n}{m\PYZus{}eq} \PY{o}{=} \PY{n}{maxwellian\PYZus{}moments}\PY{p}{(}\PY{n}{moments}\PY{p}{,} \PY{n}{dim}\PY{o}{=}\PY{l+m+mi}{2}\PY{p}{,} 
                          \PY{n}{c\PYZus{}s}\PY{o}{=}\PY{l+m+mi}{1}\PY{o}{/}\PY{n}{sqrt}\PY{p}{(}\PY{l+m+mi}{3}\PY{p}{)}\PY{p}{)}
\PY{n}{mrt} \PY{o}{=} \PY{n}{create\PYZus{}method}\PY{p}{(}\PY{n}{d2q9}\PY{p}{,} \PY{n}{moments}\PY{p}{,} \PY{err}{ω}\PY{n}{s}\PY{p}{,} 
                    \PY{n}{to\PYZus{}incompressible}\PY{p}{(}\PY{n}{m\PYZus{}eq}\PY{p}{)}\PY{p}{)}      
\end{Verbatim}

%% file: listings/output/04-ce-analysis.ipy.tex
\begin{Verbatim}[commandchars=\\\{\},fontsize=\footnotesize]
\PY{g+gp}{\PYZgt{}\PYZgt{}\PYZgt{} }\PY{n}{ce} \PY{o}{=} \PY{n}{ChapmanEnskogAnalysis}\PY{p}{(}\PY{n}{mrt}\PY{p}{)}
\PY{g+gp}{\PYZgt{}\PYZgt{}\PYZgt{} }\PY{n}{ce}\PY{o}{.}\PY{n}{get\PYZus{}bulk\PYZus{}viscosity}\PY{p}{(}\PY{p}{)}
\PY{g+go}{\PYZhy{}1/9 \PYZhy{} 1/(3*ω\PYZus{}1) + 5/(9*ω\PYZus{}0)}
\PY{g+gp}{\PYZgt{}\PYZgt{}\PYZgt{} }\PY{n}{ce}\PY{o}{.}\PY{n}{get\PYZus{}macroscopic\PYZus{}equations}\PY{p}{(}\PY{p}{)}\PY{p}{[}\PY{l+m+mi}{0}\PY{p}{]}
\PY{g+go}{∂\PYZus{}t ρ + ∂\PYZus{}0 u\PYZus{}0 + ∂\PYZus{}1 u\PYZus{}1}
\end{Verbatim}

%% file: listings/output/03-ubb.py.tex
\begin{Verbatim}[commandchars=\\\{\},fontsize=\footnotesize]
\PY{k}{def} \PY{n+nf}{vel\PYZus{}bounce\PYZus{}back}\PY{p}{(}\PY{n}{f}\PY{p}{,} \PY{n}{c}\PY{p}{,} \PY{n}{method}\PY{p}{,} \PY{n}{vel}\PY{p}{)}\PY{p}{:}
    \PY{n}{c\PYZus{}s} \PY{o}{=} \PY{n}{method}\PY{o}{.}\PY{n}{speed\PYZus{}of\PYZus{}sound}
    \PY{n}{w\PYZus{}q} \PY{o}{=} \PY{n}{method}\PY{o}{.}\PY{n}{weights}\PY{p}{[}\PY{n}{method}\PY{o}{.}\PY{n}{stencil}\PY{o}{.}\PY{n}{idx}\PY{p}{(}\PY{n}{c}\PY{p}{)}\PY{p}{]}
    \PY{n}{vel\PYZus{}term} \PY{o}{=} \PY{l+m+mi}{2} \PY{o}{/} \PY{n}{c\PYZus{}s}\PY{o}{*}\PY{o}{*}\PY{l+m+mi}{2} \PY{o}{*} \PY{n}{c} \PY{o}{*} \PY{n}{v} \PY{o}{*} \PY{n}{w\PYZus{}q}
    \PY{k}{return} \PY{n}{f}\PY{o}{.}\PY{n}{center}\PY{p}{(}\PY{n}{c}\PY{p}{)} \PY{o}{\PYZhy{}} \PY{n}{vel\PYZus{}term}
\end{Verbatim}

%% file: listings/output/02-smagorinsky.py.tex
\begin{Verbatim}[commandchars=\\\{\},fontsize=\footnotesize]
\PY{n}{S}\PY{p}{,} \PY{err}{ω} \PY{o}{=} \PY{n}{symbols}\PY{p}{(}\PY{l+s+s2}{\PYZdq{}}\PY{l+s+s2}{|S|, ω}\PY{l+s+s2}{\PYZdq{}}\PY{p}{,} \PY{n}{positive}\PY{o}{=}\PY{n+nb+bp}{True}\PY{p}{)}

\PY{n}{f\PYZus{}neq} \PY{o}{=} \PY{n}{pre\PYZus{}collision\PYZus{}symbols}\PY{p}{(}\PY{p}{)} \PY{o}{\PYZhy{}} \PY{n}{equilibrium\PYZus{}symbols}\PY{p}{(}\PY{p}{)}
\PY{err}{Π} \PY{o}{=} \PY{n}{frobenius\PYZus{}norm}\PY{p}{(}\PY{n}{second\PYZus{}order\PYZus{}moment\PYZus{}matrix}\PY{p}{(}\PY{n}{f\PYZus{}neq}\PY{p}{)}\PY{p}{)}

\PY{n}{eqs} \PY{o}{=} \PY{p}{[} \PY{n}{Eq}\PY{p}{(}\PY{err}{ω}\PY{p}{,} \PY{err}{ω}\PY{n}{\PYZus{}from\PYZus{}ν}\PY{p}{(} \PY{err}{ν}\PY{n}{\PYZus{}from\PYZus{}ω}\PY{p}{(}\PY{err}{ω}\PY{n}{\PYZus{}0}\PY{p}{)} \PY{o}{+} \PY{n}{C\PYZus{}S}\PY{o}{*}\PY{o}{*}\PY{l+m+mi}{2} \PY{o}{*} \PY{n}{S} \PY{p}{)}\PY{p}{)}\PY{p}{,}
        \PY{n}{Eq}\PY{p}{(}\PY{n}{S}\PY{p}{,} \PY{l+m+mi}{3} \PY{o}{*} \PY{err}{ω} \PY{o}{/} \PY{l+m+mi}{2} \PY{o}{*} \PY{err}{Π}\PY{p}{)} \PY{p}{]}

\PY{n}{effective\PYZus{}ω} \PY{o}{=} \PY{n}{solve}\PY{p}{(}\PY{n}{eqs}\PY{p}{,} \PY{p}{[}\PY{err}{ω}\PY{p}{,} \PY{n}{S}\PY{p}{]}\PY{p}{)}\PY{p}{[}\PY{err}{ω}\PY{p}{]}
\end{Verbatim}